# Laser-Induced Rashba Spin-Orbit Torques in Multiferroic Semiconductor (Ge,Mn)Te

*Zeynab Sadeghi, Tomáš Ostatnický, Eva Schmoranzerová, Jozef Kimák, Dominik Kriegner, Helena Reichlová, Lukáš Nádvorník, Gunther Springholz, J. Hugo Dil, Juraj Krempasky, and Petr Němec**

Z. Sadeghi, T. Ostatnický, E. Schmoranzerová, J. Kimák, L. Nádvorník, P. Němec
Faculty of Mathematics and Physics, Charles University, Prague, Czech Republic
* E-mail: petr.nemec@matfyz.cuni.cz
D. Kriegner, H. Reichlová
Institute of Physics ASCR, v.v.i., Prague 6, Czech Republic
G. Sprinhgholtz
Institut für Halbleiter-und Festkörperphysik, Johannes Kepler Universität, Linz, Austria
J. H. Dil
Institut de Physique, École Polytechnique Fédérale de Lausanne, Lausanne, Switzerland
J. H. Dil, J. Krempasky
Centre for Photon Science, Paul Scherrer Institut, Villigen, Switzerland.

**The multiferroic semiconductor (Ge,Mn)Te exhibits ferroelectricity, strong Rashba spin-orbit coupling, and carrier-mediated magnetic order, making it a unique platform for exploring the interplay between electronic, structural, and magnetic degrees of freedom. In this study, we investigate the ultrafast magnetization dynamics of $Ge_{0.85}Mn_{0.15}Te$ using time-resolved magneto-optical spectroscopy. By separating magnetic and nonmagnetic contributions in the transient response, we identify two distinct laser-induced magnetic phenomena, both arising from a photoinduced effective spin-orbit torque. Coherent magnetization precession is driven by a laser-induced increase in hole concentration, which modifies the occupation of Rashba-split, spin-locked valence band states and shifts the position of the magnetic easy axis. The transient change in magnetic properties, observed as variations in the amplitude and coercive field of the pump-induced change in ellipticity hysteresis loops, is attributed to laser-induced modifications of the ferroelectric sublattice displacement, which affect the ferroelectric polarization and the associated Rashba spin-orbit interaction. While the first type of optical spin-orbit torque has already been observed in the diluted magnetic semiconductor (Ga,Mn)As, the second effect is unique to the multiferroic Rashba semiconductor (Ge,Mn)Te, opening new opportunities for all-optical manipulation of magnetic order and spin-orbit torque generation in spin-orbitronic devices.**

## 1. Introduction

Efficient manipulation of magnetization through spin–orbit coupling (SOC) is a central goal of modern spintronics. In solids lacking inversion symmetry, SOC lifts the spin degeneracy of electronic states and creates momentum-dependent spin textures via the Rashba and Dresselhaus effects. The Rashba effect is typically much stronger and therefore more attractive for practical device applications but it requires specific inversion symmetry breaking to influence bulk electron states. The Rashba spin–orbit interaction enables efficient interconversion between charge and spin degrees of freedom and underlies phenomena such as the Edelstein effect, inverse Edelstein effect, and current-induced spin–orbit torques (SOTs).[1,2] Unlike conventional spin-transfer torques,[3,4] which require spin-polarized currents injected from a magnetic reference layer, SOTs originate directly from spin–orbit interactions and thus provide an efficient method for electrical control of magnetization.[5,6] As a result, materials with strong Rashba coupling have attracted significant attention as promising platforms for low-power memory technologies, spin–orbitronic devices, and ultrafast information processing.[1,2]

Among the various Rashba systems, ferroelectric Rashba semiconductors are a particularly attractive class of materials because inversion symmetry breaking originates from a switchable ferroelectric polarization. The prototypical member of this family is $\alpha$-GeTe, a semiconductor with a 0.85 eV indirect band gap[7] that undergoes a rhombohedral distortion below its ferroelectric transition temperature near 700 K.[8] The relative displacement of Ge and Te sublattices along the [111] crystallographic direction generates a robust spontaneous polarization and produces giant Rashba splitting of both bulk and surface electronic states.[9-11] Room-temperature ferroelectricity, an exceptionally large Rashba parameter $\alpha_R$ of around 4 eVÅ,[9-11] and semiconducting character have established GeTe as a model system for investigating Rashba physics and ferroelectric spintronics. The coupling between ferroelectricity and Rashba spin splitting has stimulated extensive efforts to control its electronic structure. Electrical switching of the ferroelectric polarization has been shown to reverse the spin texture, highlighting the potential of GeTe for nonvolatile spintronic applications.[12,13] More recently, ultrafast transient enhancement of the Rashba parameter by femtosecond laser pulses was demonstrated.[14] This experiment revealed that Rashba spin splitting can be dynamically controlled on sub-picosecond timescales and established a direct link between ultrafast optical excitation, ferroelectric polarization, and Rashba spin-orbit coupling. At the same time, this raises the intriguing question of whether such photoinduced

Rashba modulation could be harnessed to manipulate magnetic order in systems where Rashba coupling is intrinsically linked to magnetization.

A natural platform for addressing this question is $Ge_{1-x}Mn_xTe$, the magnetic counterpart of GeTe. In this material, magnetic doping preserves the polar crystal structure of the parent compound while introducing carrier-mediated magnetic order, resulting in a multiferroic material that simultaneously exhibits ferroelectricity, giant Rashba spin splitting, and broken time-reversal symmetry. The coexistence of inversion symmetry breaking and magnetism gives rise to a Rashba–Zeeman electronic structure characterized by strong spin-momentum locking, exchange splitting, and enhanced Berry curvature effects.[12,14] Notably, current-induced magnetization switching mediated by the bulk Rashba–Edelstein effect has been experimentally demonstrated, providing direct evidence that non-equilibrium spin polarization originating in Rashba effect can efficiently manipulate magnetic order in this material.[15] Despite significant progress in understanding the static electronic structure and transport properties of $Ge_{1-x}Mn_xTe$, the nonequilibrium dynamics of its Rashba-coupled magnetic state remain largely unexplored. In particular, it remains unknown whether photoinduced modifications of the Rashba interaction[14,16,17] can, through the coupling between the two ferroic orders, generate effective spin-orbit torques capable of driving coherent magnetization dynamics. Addressing this question requires both ultrafast optical control and a reliable means of disentangling the magnetic and nonmagnetic contributions to the transient optical response of $Ge_{1-x}Mn_xTe$.

In this paper, we investigate the ultrafast magnetization dynamics of $Ge_{0.85}Mn_{0.15}Te$ using time-resolved magneto-optical spectroscopy. By carefully analyzing the signals following femtosecond optical excitation, we uncover complex magnetization dynamics, including magnetization precession and time-dependent changes in the measured hysteresis loops. Our experimental data cannot be explained by a simple laser-induced increase in sample temperature. Instead, they are consistent with a scenario in which laser pulses modify the magnetic state of the system on ultrashort time scales by altering the ferroelectric polarization, which, through Rashba coupling, induces changes in the magnetic state. This represents an important step toward all-optical spin-orbitronics, in which ultrafast optical control of Rashba coupling is directly converted into magnetic control. In a broader context, our work describes a new type of nonthermal photomagnetic excitation mechanism[18] that can be used for ultrafast manipulation of magnetization in a specific class of multiferroic materials.

## 2. Results

### 2.1. Samples and experiment

An epitaxial $Ge_{1-x}Mn_xTe$ film with nominal $x_{Mn}$ of 0.15 and a thickness of 50 nm (henceforth GeMnTe sample), grown on a $BaF_2$(111) substrate by molecular beam epitaxy, was studied. At this concentration, magnetic ion (Mn) doping induces hole-mediated long-range magnetic ordering with the corresponding ordering temperature $T_C$ up to 200 K,[19-21] while preserving the polar crystal structure of the parent $\alpha$-GeTe compound, which is responsible for the ferroelectric ordering.[21,22] As a reference, we used a 50 nm film with $x_{Mn}$ of 0.30 (henceforth reference $Ge_{0.70}Mn_{0.30}Te$ sample) in which the ferroelectric ordering is significantly reduced by Mn doping.[21,20] As the pump-probe results obtained for this reference sample do not show any signals attributable to spin-orbit torque, they are shown only in the Supplementary Information (see Supplementary Figure S4). For more details on sample preparation and characterization, see Methods and Supplementary Note 1.

The dynamical changes induced in GeMnTe sample by femtosecond laser pulses at 15 K were investigated using the magneto-optical (MO) pump-probe technique in a transmission geometry,[23] as schematically shown in **Figure 1a**. Absorption of the pump pulses causes a change in the sample's MO activity, which is measured by a time-delayed linearly polarized probe. (For further details, see Methods.) Because the experiment is performed in transmission geometry, the measured signals primarily reflect the properties of bulk Rashba-split electronic states and their coupling to the magnetic order. As shown below, the measured time-resolved signals in GeMnTe sample depend on many experimental parameters, as effects related not only to magnetic ordering contribute to the observed probe-polarization changes.[24] Consequently, to disentangle the magnetization-related components of the signal, we measured the transient MO signals both as a function of the pump–probe time delay, as is conventional in these experiments, and as a function of the magnetic field at a fixed time delay.

In ferroelectric materials, inversion symmetry is inherently broken by the polarization $P$, which arises from the relative displacement of the anion and cation sublattices (sublattice shift in the following). In $\alpha$-GeTe, the pronounced rhombohedral lattice distortion, which causes cation and anion shift of up to 10% of the (111) lattice plane spacing (see Figure 1b), leads to a giant spin splitting with a Rashba parameter $\alpha_R$ as large as 2–4 eVÅ.[9-11] Ferroelectricity is weakened by Mn doping because the Mn ion displacement is smaller than that of Ge, but it can persist for Mn concentrations up to 50%.[21,25] The absorption of a strong femtosecond laser pulse changes the distance between Ge/Mn and Te atoms,[14,16,17] as schematically shown in

Figure 1c. The collinear alignment of electric and magnetic polarization in GeMnTe leads to the opening of a Zeeman gap around the Dirac point of the Rashba bands,[12,15,25] as schematically depicted in Figure 1d. Therefore, the eventual laser-induced change in Rashba spin splitting can significantly alter the resulting band structure, which should, in turn, affect the magnetic system through an effective spin-orbit torque. As described in detail below, this scenario is consistent with the experimental data shown in Figure 1e.

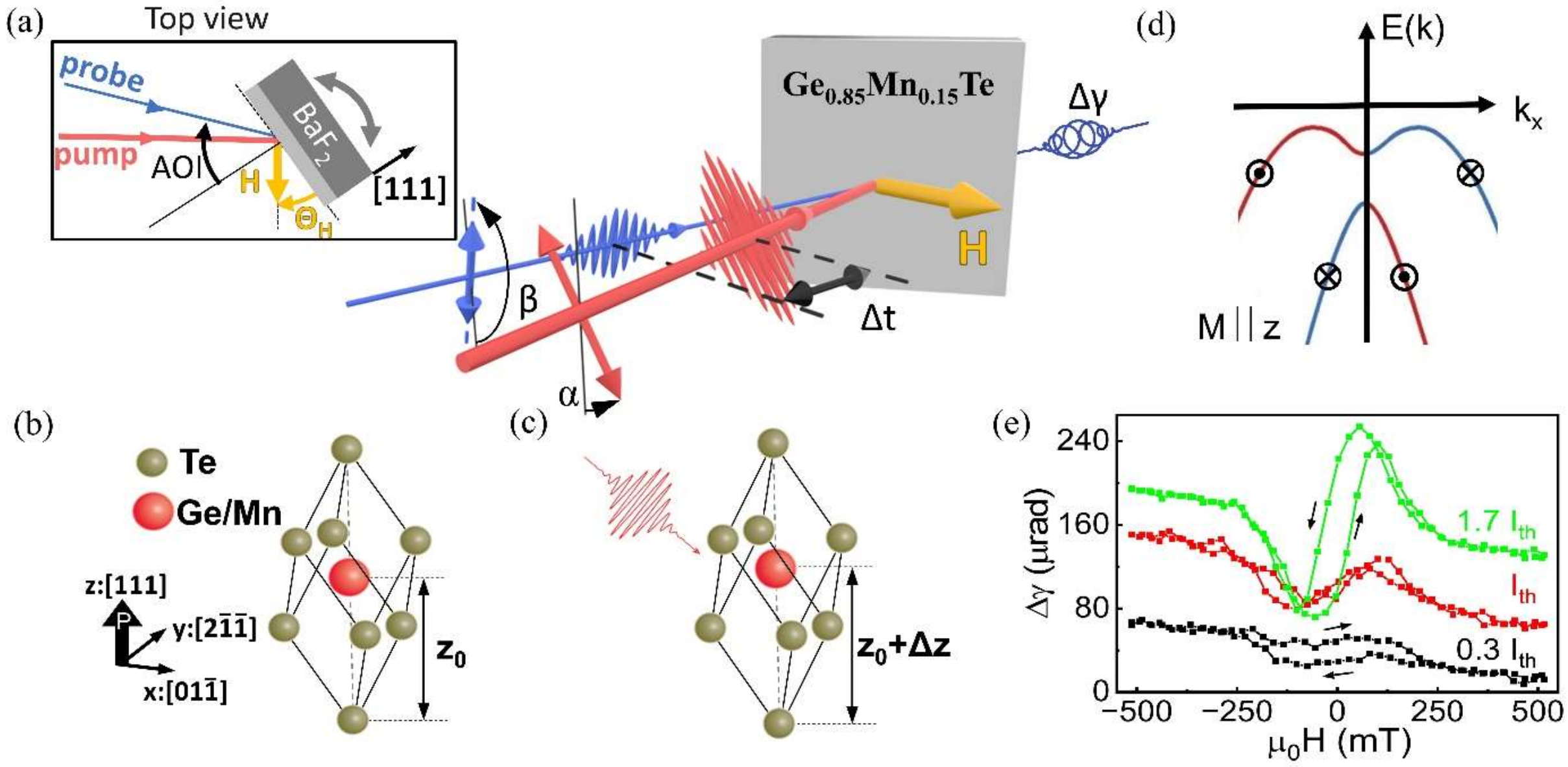

**Figure 1.** Laser-induced Rashba spin-orbit torque in (Ge,Mn)Te. **a**, Schematic of the experimental pump-probe setup used to induce and detect magnetic dynamics. The orientations of the linear polarization of the pump and probe pulses are described by angles $\alpha$ and $\beta$, respectively. The pump-induced change in ellipticity ($\Delta\gamma$) of the transmitted probe pulses is measured as a function of the time delay ($\Delta t$) between the pump and probe pulses. Inset: Top view of the setup. The sample can be tilted around the axis perpendicular to the depicted plane, resulting in a change in the probe beam angle of incidence (*AOI*) and the orientation of the magnetic field ($H$) relative to the sample plane ($\Theta_H$); depicted angles correspond to positive values. The pump beam propagates perpendicular to $H$. **b**, Rhombohedrally distorted $\alpha$-GeTe unit cell with Mn atoms substituting for a fraction of Ge atoms. The indicated crystallographic directions are given with respect to the cubic $BaF_2$ substrate and coincide with the corresponding directions in the primitive two-atom rhombohedral unit cell of GeTe. **c**, The displacement of atoms along the [111] crystallographic direction, responsible for the ferroelectricity described by polarization ($P$, black arrow), is altered by absorption of a femtosecond laser pulse. **d**, Rashba-type spin-polarized valence bands with opposite spin momenta and a gap opening due to time-reversal symmetry breaking by magnetic order when the magnetization ($M$) points along the out-of-plane [111] crystallographic direction. **e**, Pump-induced change in hysteresis loops measured for different pump intensities at $\Delta t$ = 20 ps. Note the pronounced change in the character of the hysteresis loop below and above the threshold fluence $I_{th} \approx 6.5$ mJ cm$^{-2}$; arrows indicate the direction of the magnetic field change. Pump wavelength 1 000 nm, probe wavelength 670 nm, AOI = 0°, $\Theta_H$ = -10°.

## 2.2. Static magneto-optical characterization

Mn doping of GeTe produces not only substitutional but also two types of interstitial positions for Mn atoms.[26] The direct antiferromagnetic interaction between Mn interstitials, along with the hole-mediated ferromagnetic Ruderman–Kittel–Kasuya–Yosida (RKKY) interaction between substitutional Mn atoms, leads to frustration and competition. Experimentally, magnetization curves reach saturation for fields above 5 T, so the system was originally regarded as exhibiting long-range ferrimagnetic order, but due to the coexistence of substitutional and interstitial Mn it turned out to be better described as an inhomogeneous admixture of (Mn/hole-rich) ferrimagnetic clusters within a paramagnetic (Mn/hole-pure) environment.[27-30] Finally, it was concluded that GeMnTe is a correlated spin glass at low temperatures (see **Figure 2a**), where coupling between individual ferrimagnetic clusters provides the long-range order.[26]

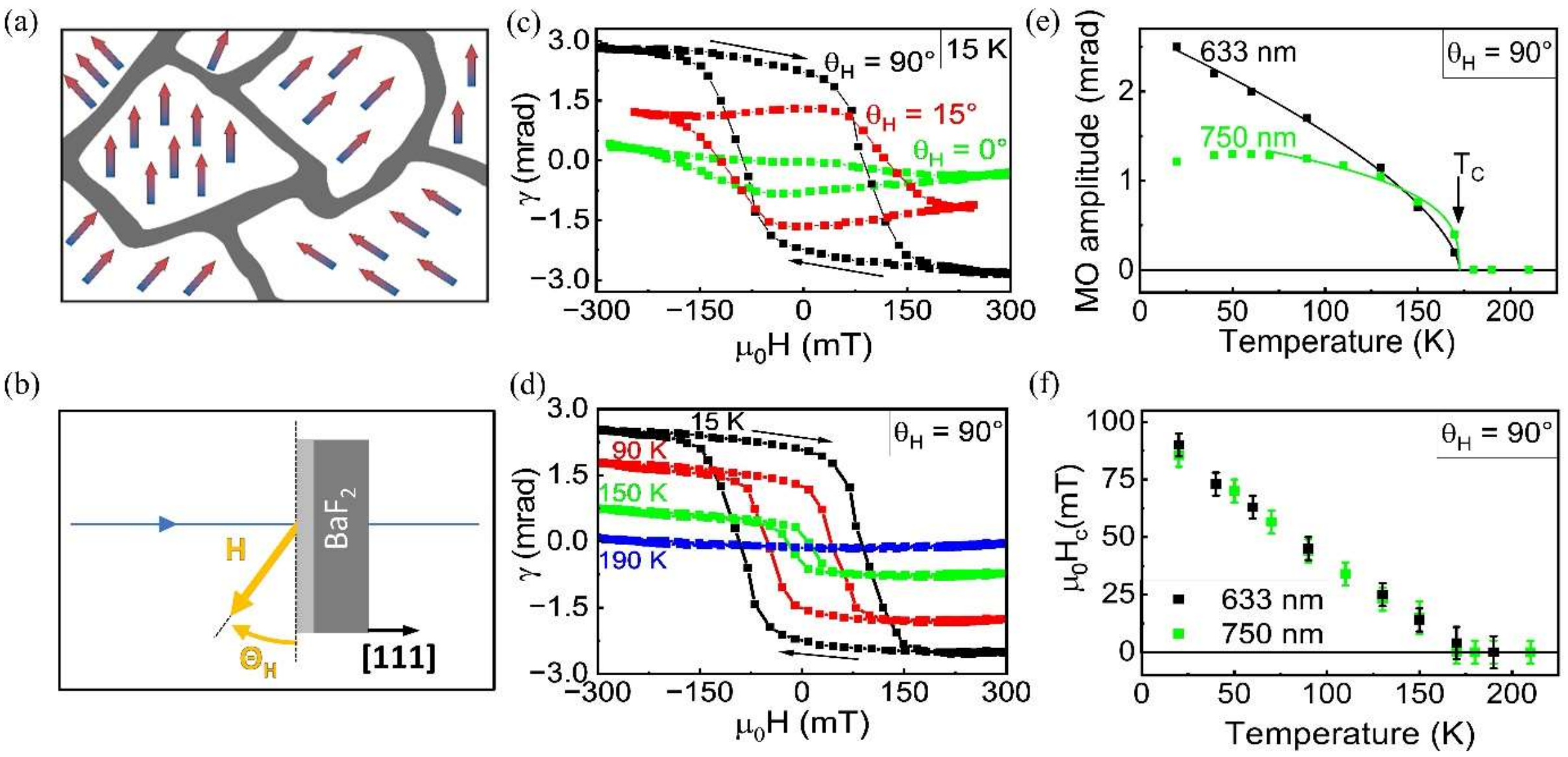

**Figure 2.** Static magneto-optical (MO) characterization of GeMnTe sample. **a**, Pictogram showing the assumed correlated spin glass state of GeMnTe sample at low temperatures. **b**, Schematic of the experimental setup with the laser beam incident perpendicular to the sample and the magnetic field applied at different angles $\Theta_H$ relative to the sample plane. **c**, Change in light ellipticity ($\gamma$) measured at 633 nm as a function of magnetic field magnitude ($H$) for field directions $\Theta_H$ = 90° (out-of-plane), 15°, and 0° (in-plane). **d**, Temperature dependence of hysteresis loops for $\Theta_H$ = 90° measured at 633 nm. **e**,**f**, Temperature dependence of MO amplitudes (e), corresponding to half the loop height, and coercive fields (f) extracted from the hysteresis loops measured for $\Theta_H$ = 90° at 633 nm and 750 nm (points). Lines are fits to the formula $MO \sim (T_c - T)^{\beta}$ with $T_C$ = 172 ± 5 K (the error includes both the fitting error and the uncertainty in the sample temperature in the cryostat). Note that spectral dependence of MO signal is shown in Figure 5f.

As a prerequisite for the pump-probe experiments, we performed a static MO characterization of the GeMnTe sample. In the experimental geometry used (Figure 2b), with the laser beam perpendicular to the sample plane, the measured MO signal corresponds solely to the Faraday effect (see Methods), whose magnitude is proportional to the out-of-plane projection of the magnetization. The low-temperature data shown in Figure 2c reveal the presence of remanent magnetization, whose out-of-plane magnitude strongly depends on the magnetization history. In particular, the MO signals measured at zero applied field differ considerably for hysteresis loops recorded with fields applied perpendicular to the sample plane (black curve in Figure 2c) or within the sample plane (green curve in Figure 2c), which is a feature expected for a clustered spin glass. (See also Supplementary Figure S1 for corresponding SQUID data, which show hysteretic behavior for both out-of-plane and in-plane magnetic fields.) As the temperature increases (Figure 2d,2e, and 2f), the amplitudes of the hysteresis loops and the coercive fields decrease monotonically, revealing the sample's ordering temperature, $T_C = 172 \pm 5$ K. Importantly, the sign of the hysteresis loop remains unchanged for all temperatures and magnetic-field angles, positive at −300 mT and negative at +300 mT.

## 2.3. Ultrafast pump-probe experiment

### *2.3.1. Disentangling magnetic signals from measured time-resolved responses*

The goal of our magneto-optical (MO) pump-probe experiment is to detect changes in the MO response caused by a strong (pump) laser pulse in the studied sample, from which laser-induced changes in magnetic ordering and its dynamics can be deduced. To achieve this, the pump-induced change in the much weaker (probe) pulse ellipticity ($\Delta\gamma$) and/or rotation ($\Delta\beta$) is measured as a function of the pump–probe time delay ($\Delta t$). However, the sample's MO response is not the only effect that can influence the probe polarization; strain and pump-induced lattice heating, which modify the sample's complex refractive index, also cause probe polarization changes that are purely nonmagnetic in origin.[24] These effects are rather strong in GeMnTe, which is likely why no pump-probe experimental results have been reported for this material until now. To address this, we measured not only the pump-induced probe polarization changes as a function of the pump–probe time delay but also the pump-induced changes in hysteresis loops and in sample transmission. The former directly reveals the change in magnetic ordering in the material, while the latter provides information about the dynamics of non-equilibrium carriers photoexcited in the material by the pump pulse. All time-resolved measurements were performed at a temperature of 15 K.

In **Figure 3a**, we show the effect of pump polarization on hysteresis loops measured in the experimental geometry depicted in Figure 1a. The pump-induced transmission change is independent of both the applied magnetic field (not shown) and the pump polarization (see Figure 3b), indicating that the number of photoexcited carriers responsible for this signal does not depend on pump polarization. In contrast, the pump-induced change in probe ellipticity dynamics depends strongly on both the applied field and the pump polarization (Figure 3c and 3d). To determine, which part of the measured signal is sensitive to the laser-induced change in magnetic properties in the material, we plot in Figure 3e and 3f the dynamics of the component $\Delta\gamma^{dif}$, corresponding to the difference between measured signals for opposite field polarities, and $\Delta\gamma^{aver}$, corresponding to their average (see Methods for their definitions). Figure 3e shows that, regardless of pump polarization, absorption of pump pulses induces magnetization precession in the GeMnTe sample, as also observed in the canonical diluted magnetic semiconductor (Ga,MnAs).[31,32] However, in the GeMnTe sample we also observe strong pump-polarization-dependent signals (Figure 3f). A complementary perspective on these time-resolved data is provided by measuring the pump-induced change in hysteresis loops for different pump polarizations (Figure 3a). These data show that pump pulses of any polarization, both linear and circular, alter the hysteresis loop in the same way, but produce a magnetic-field-independent background signal (see also Supplementary Figure S2a and S2b). Importantly, our experiments revealed (Supplementary Figure S3) that while $\Delta\gamma^{dif}$ data were strongly suppressed at temperatures above the sample Curie temperature, the $\Delta\gamma^{aver}$ data were not significantly affected by the temperature increase. Presumably, this signal component can be partially related to the modification of the polarization vector $P$ in the $Ge_{0.85}Mn_{0.15}Te$ sample,[33] as the characteristic transition temperature in GeTe is as high as 700 K.[8] However, the experiments in the reference $Ge_{0.70}Mn_{0.30}Te$ sample, where ferroelectric ordering is significantly reduced by high Mn doping, revealed similar $\Delta\gamma^{aver}$ signals. Therefore, this component is likely a mixture of signals with different origins, whose disentanglement is not straightforward.

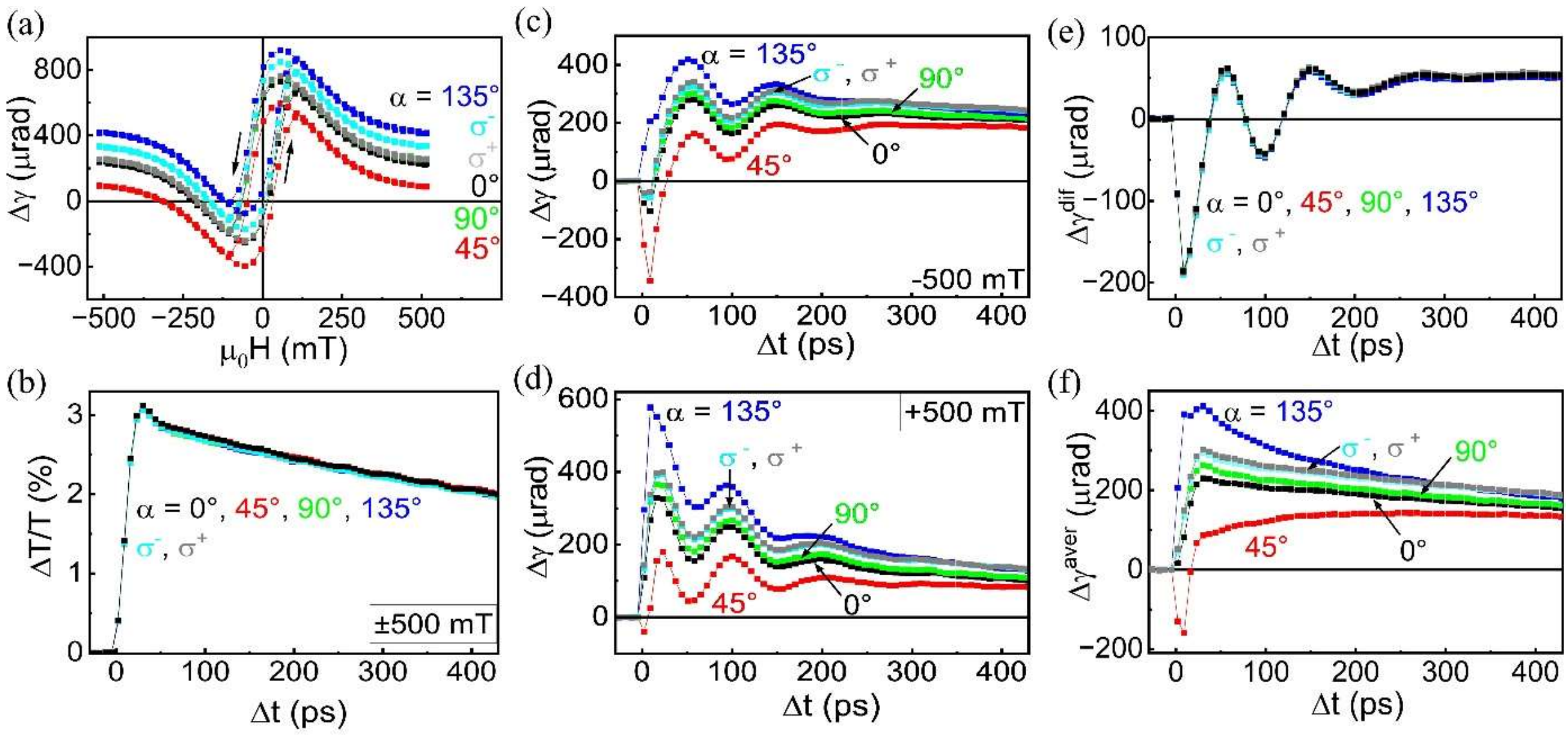

**Figure 3.** Influence of pump polarization. **a**, Pump-induced changes in hysteresis loops measured at $\Delta t$ = 20 ps for different pump linear polarizations $\alpha$ and circular polarizations $\sigma$; arrows indicate the direction of the magnetic field change. **b**, Dynamics of transient transmission $\Delta T/T$. **c**,**d**, Dynamics of ellipticity measured at -500 mT (c) and +500 mT (d). **e**,**f**, Dynamics of components $\Delta\gamma^{dif}$ (e) and $\Delta\gamma^{aver}$ (f), see Methods for their definitions. Pump fluence 1.3 $I_{th}$, wavelength 820 nm; probe wavelength 633 nm, polarization $\beta$ = 0°; AOI = -5°, $\Theta_H$ = -15°.

In **Figure 4**, we examine the role of probe polarization in the measured signals. Similar to the pump polarization dependence, the measured MO signals depend strongly on the probe light polarization (Figure 4a, 4c, and 4d), whereas the transmission change (Figure 4b) does not. Probe polarization dependence is, in principle, a powerful tool for separating different MO signals.[34] Ideally, when MO effects sensitive to both in-plane and out-of-plane projections of magnetization have comparable magnitudes – which depend strongly on the probe wavelength – it is even possible to fully reconstruct the three-dimensional magnetization trajectory after the laser pulse impacts the studied material.[32,35]

The data measured in the GeMnTe sample at a probe wavelength of 633 nm show that the signal sensitive to magnetization precession (Figure 4e) does not depend on probe polarization, which is typical of a Faraday MO signal. However, closer inspection of the measured hysteresis loops reveals that the Voigt MO effect, which can be separated from the data by subtracting measurements obtained for two orthogonal probe polarizations,[32] also contributes to the measured signal (see Supplementary Figure S2d). Overall, the signal components $\Delta\gamma^{dif}$ and $\Delta\gamma^{aver}$ can be thus interpreted as follows. The signal $\Delta\gamma^{dif}$, which changes sign when the field direction is reversed, corresponds to odd-in-magnetization MO signals (such as Faraday and Kerr effects). The signal $\Delta\gamma^{aver}$, which does not change sign with the field direction, represents signals not connected with magnetic ordering in the sample and signals corresponding to even-in-magnetization MO effects (such as the Voigt effect).

In summary, identifying the $\Delta\gamma^{dif}$ signal as the component that carries information about the laser-induced magnetization dynamics forms the basis of our subsequent analysis. This identification allows us to isolate the ultrafast evolution of the magnetic state in the GeMnTe sample and systematically investigate its dependence on pump–probe time delay, magnetic field and excitation fluence, as discussed below.

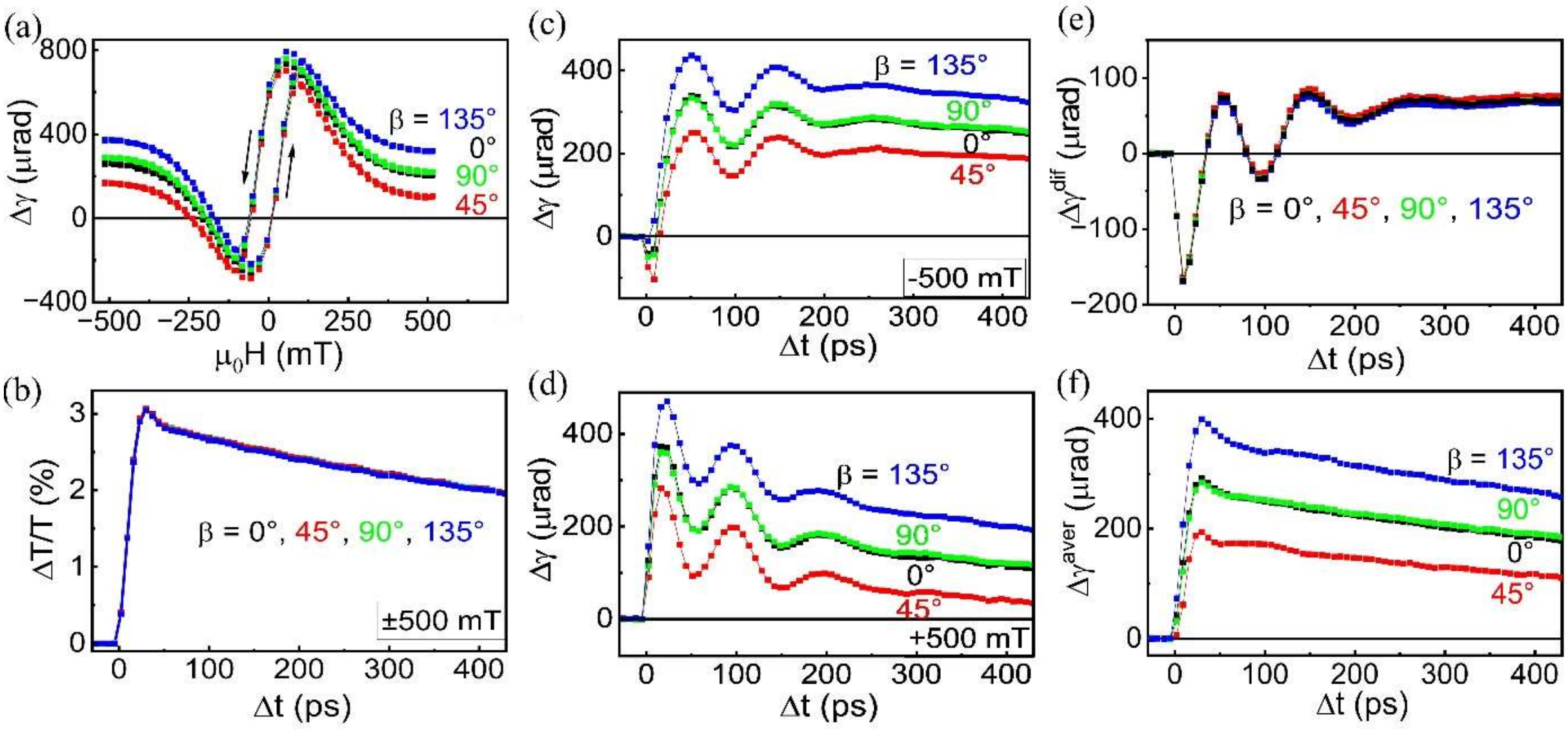

**Figure 4.** Influence of probe polarization. **a**, Pump-induced changes in hysteresis loops measured at $\Delta t$ = 20 ps for different orientations of probe linear polarization $\beta$; the arrows indicate the direction of the magnetic field change. **b**, Dynamics of transient transmission $\Delta T/T$. **c**,**d**, Dynamics of ellipticity measured at -500 mT (c) and +500 mT (d). **e**,**f**, Dynamics of components $\Delta\gamma^{dif}$ (e) and $\Delta\gamma^{aver}$ (f). Pump fluence 1.3 $I_{th}$, wavelength 820 nm, circular polarization $\sigma$; probe wavelength 633 nm; AOI = -5°, $\Theta_H$ = -15°.

*2.3.2. Laser-induced spin-orbit torque*

In the previous section, we showed that the complex optical response of GeMnTe sample to laser pulses can be significantly simplified if the magnetic-field-dependent signal component $\Delta\gamma^{dif}$ is extracted from the measured data. Furthermore, with an appropriate choice of pump and probe polarizations, this signal can already dominate in the raw data. To achieve this condition, circularly polarized pump pulses (see Supplementary Figure S2a) and probe pulses with $\beta$ = 0° (Supplementary Figure S2c) were used in the following experiments, revealing magnetization dynamics that do not depend on the pump polarization (Figure 3e). In principle, such signals can result from either a pump-induced temperature rise or an increase in carrier concentration, which can be distinguished by the sign of the MO signals (see also the Discussion section). A pump-induced temperature increase leads to demagnetization of the ferromagnet, resulting in a reduction of its MO response.[18] Consequently, if this were the dominant signal, the static (Figure 2d) and low-intensity dynamical data (Figure 1e) would have opposite signs, which is not the case in our experiment. Moreover, as discussed in detail below, a nontrivial intensity dependence of the measured MO signals was observed, revealing pronounced changes in the hysteresis loops above the threshold pump fluence $I_{th}$, in contrast to the smooth temperature evolution of the static MO response (Figure 2d–f). On the other hand, the measured data are fully consistent with the expected laser-induced transient increase in hole concentration and Rashba-mediated changes in the GeMnTe band structure, which affect the sample's magnetic ordering through an effective spin-orbit torque.

In **Figure 5a**, we show the time evolution of pump-induced changes in the hysteresis loop for a pump fluence of 1.3 $I_{th}$. At this intensity, the pump-induced hysteresis loops exhibit substantially different coercive fields of 35 ± 5 mT (inset in Figure 5d) compared with the static case, 120 ± 5 mT (data for $\Theta_H$ = 15° in Figure 2c), at the same base temperature. Moreover, the sign of the pump-induced hysteresis loop is reversed, and the loop has a qualitatively different shape. As shown in Figure 5a, the magnetic-field dependence is nonmonotonic: the signal first decreases and then increases even after saturation (closing of the hysteresis loop). Additionally, as shown in Figure 1e, the data measured for 1.7 $I_{th}$ reveal that while the dominant magnetization change occurs at fields of ≈ 90 mT, complete saturation is reached only for fields approaching ≈ 250 mT. This behavior indicates a rather complex change in the magnetic anisotropy landscape induced by the pump pulse.

The dynamics of transient transmission is shown in Figure 5b, and the dynamics of the field-direction-dependent MO signal components at 500 mT and 16 mT are shown in Figure 5c and Figure 5e, respectively (the complete set of time-resolved MO data within one hysteresis-loop branch is provided in Supplementary Figure S5). The most notable feature is the very fast rise of the MO signal for both 500 mT (see inset in Figure 5c) and 16 mT (inset in Figure 5e), indicating that the pump-induced change in the magnetic properties of the GeMnTe sample is completed in less than ≈ 1 ps after absorption of the pump pulse (see Supplementary Note 2.8 for a detailed discussion). Interestingly, the signal at 500 mT changes sign within approximately 50 ps (Figure 5c), whereas the signal at 16 mT does not (Figure 5e). On longer time scales, both signals decay toward equilibrium (Figure 5d) at a rate similar to that of the transient transmission (Figure 5b). The oscillatory MO signal corresponds to pump-induced precession of magnetization, which can be fitted by Equation (3) in Methods.[31,36] (See Supplementary Figure S6 for precession signals measured at several magnetic fields and fitted parameters.) This signal represents an optical analogue of ferromagnetic resonance (FMR), from which many micromagnetic parameters can be deduced.[37] However, such detailed analysis is beyond the scope of this paper. Here, we note that the measured MO data can be fitted well by Equation 3 for time delays as short as 2 ps (see inset in Figure 5c).[38] This again demonstrates that even at picosecond timescales, the precession of magnetization is already fully developed, indicating a sub-picosecond duration of the torque.

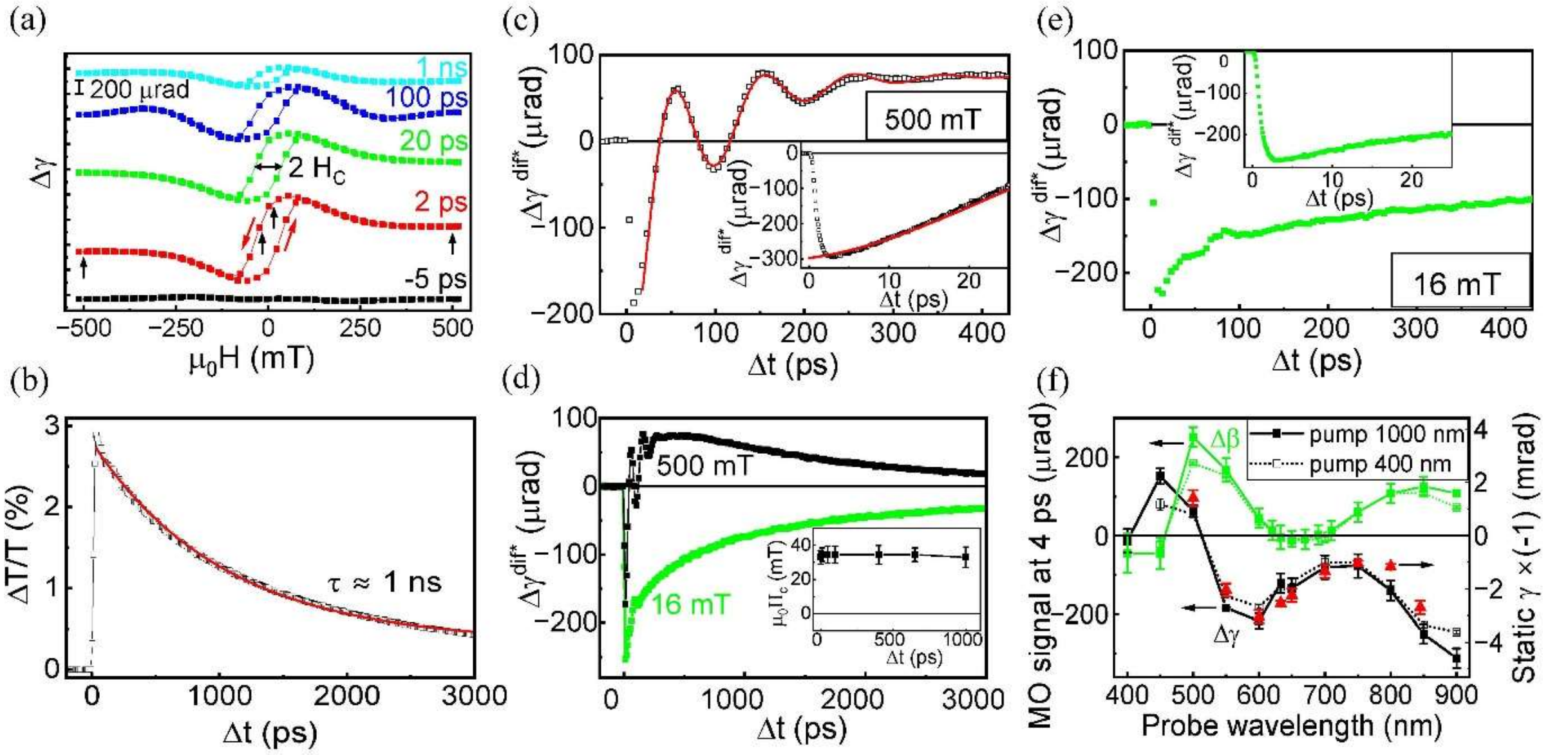

**Figure 5.** Dynamics of the measured signals for pump fluence above $I_{th}$. **a**, Pump-induced changes in hysteresis loops measured at the indicated time delays; the data are vertically shifted for clarity. The oblique arrows indicate the field direction changes, the horizontal double arrow illustrates the coercive field $H_c$, and the vertical arrows indicate the fields where time-resolved measurements, used to compute $\Delta\gamma^{dif*}$ depicted in parts c and d, were measured. **b**, Dynamics of transient transmission $\Delta T/T$ (points); the red solid line is a fit using a monoexponential decay function plus background with the shown time constant. **c**, Dynamics of $\Delta\gamma^{dif*}$, computed from signals measured within one hysteresis loop branch, for 500 mT (points); the red solid line is a fit using Equation 3. Inset: Detail of the dynamics. **d**, Nanosecond dynamics of $\Delta\gamma^{dif*}$ for 500 mT and 16 mT. Inset: Time evolution of the coercive field deduced from the data shown in **a**. **e**, Dynamics of $\Delta\gamma^{dif*}$ for 16 mT. **f**, Comparison of MO spectra for the laser-modified and equilibrium magnetic states. Squares represent the probe-wavelength dependence of ellipticity $\Delta\gamma^{dif*}$ and rotation $\Delta\beta^{dif*}$ at $\Delta t$ = 4 ps induced for 16 mT by pump pulses at 1 000 nm and 400 nm. Triangles show the spectral dependence of the static ellipticity multiplied by (-1). a-e: Pump fluence 1.3 $I_{th}$, wavelength 820 nm, circular polarization $\sigma$; probe wavelength 633 nm, polarization $\beta$ = 0°; AOI = -5°, $\Theta_H$ = -15°. f: Pump fluence at different excitation wavelengths was adjusted to correspond to 1.3 $I_{th}$ at 820 nm (as monitored by the resulting $\Delta T/T$ signal), circular polarization $\sigma$; probe polarization $\beta$ = 0°; AOI = 0°, $\Theta_H$ = -10°.

The observation that the pump pulse significantly alters the sample's magnetic anisotropy landscape, as evidenced by changes in the hysteresis loops, raises the question of whether the laser-modified magnetic state can truly be considered a perturbed equilibrium state, or whether ultrafast laser excitation instead drives the material's magnetic state beyond the perturbative regime. To address this question, we measured the spectral dependence of the pump-induced MO signals by tuning the wavelength of the probe pulses, effectively performing magneto-optical spectroscopy on this transient magnetic state. As shown in Figure 5f, the ellipticity spectra for the pump-induced (black squares) and static (red triangles) cases are virtually identical, spectroscopically confirming that the magnetic phase is the same in both experiments, i.e., that the pump pulses do not induce a magnetic phase transition. Consequently, we propose that the pump pulses solely reconfigure the correlated spin-glass structure of the GeMnTe sample into a state with lower free energy, resulting in reduced coercive fields. Moreover, the measured spectral dependence of the MO signals is not significantly affected by the pump pulse wavelength (open and solid squares in Figure 5f; see also Supplementary Figure S8 for data measured at additional excitation wavelengths). This indicates that the exact spectral position of the photoinjected carriers within the semiconductor band structure does not play a significant role in this effect, ruling out ultrafast phonon emission during the initial cooling of photoinjected carriers as the dominant excitation mechanism (see also the Discussion section).

While the presence of ultrafast changes in the sample's magnetic properties is robust in our experiment, the exact shape of the measured MO signals is strongly influenced by the experimental geometry. In particular, as shown in detail in Supplementary Figure S7, changing the angle of incidence (AOI) of the probe pulses, which simultaneously alters the angle $\Theta_H$ between the field direction and the sample surface (see inset in Figure 1a), significantly affects the signal magnitude and dynamics. The changes in the hysteresis loops (Supplementary Figure S7d) agree with the results of static MO characterization of the GeMnTe sample (Figure 2c), where the remanent MO signal strongly depends on the direction of the applied field, a feature expected for a clustered spin glass. However, from an MO standpoint, there is a significant geometric difference between these two experiments. All static MO measurements were performed with normal incidence of light on the sample (AOI = 0°) for all directions of the applied field (see Figure 2b), a configuration that does not allow MO signals due to the longitudinal Kerr-like effect in the transmission geometry used, so only the MO Faraday effect contributes to the $\Delta\gamma^{dif}$ signal. In contrast, in a standard time-resolved experiment, a change in $\Theta_H$ is inevitably accompanied by a change in the probe beam AOI (see Figure 1a), leading to a mixture of MO signals of different origins. Specifically, while the Faraday effect is sensitive to the out-of-plane projection of the magnetization, longitudinal Kerr-like signals are sensitive to its in-plane projection. This significantly complicates the detailed interpretation of the origin of the very fast rise of the MO signal, which appears in less than 1 ps, as shown in the inset of Figure 5c, and disappears completely at certain experimental geometries (see the data measured for AOI = 0° in Supplementary Figure S7e). To address this, we constructed a modified experimental pump–probe setup, shown in Figure 6a, in which a concentric geometry of co-propagating pump and probe pulses allows AOI and $\Theta_H$ to be set to 0° simultaneously (see Methods), so that only the Faraday effect contributes to the $\Delta\gamma^{dif}$ signal.

*2.3.3. Experiment with concentric pump and probe beams*

In **Figure 6** and 7, we present experimental data measured using the concentric setup shown in Figure 6a. For this experiment, a probe wavelength of 532 nm was selected because the magnitudes of the ellipticity $\Delta\gamma$ and rotation $\Delta\beta$ magneto-optical signals are comparable at this wavelength (see Figure 5f), and therefore both signal types can be measured reliably.[40] For different pump intensities, no significant changes in the transmission recovery dynamics are observed (Figure 6b), whereas the measured MO signals change considerably. The pump-induced hysteresis loops obtained at several pump intensities are shown in Figure 6c. Aside from negative nonmagnetic background signals, these data exhibit trends very similar to those observed in the non-collinear geometry at 670 nm, as shown in Figure 1e. Specifically, the hysteresis loops at low excitation intensities closely resemble those measured in the sample's equilibrium state (Figure 2). However, when the intensity exceeds the threshold fluence $I_{th} \approx 6.5$ mJ cm$^{-2}$, the shape of the loops changes (see also Supplementary Figure S12 where more data are shown). Because the pump-induced transmission change, which is proportional to the number of photoexcited carriers, follows the expected linear intensity dependence (green points in Figure 6d), saturation of sample absorption across the entire intensity range studied is ruled out. Despite this, the amplitude of the hysteresis loops changes non-monotonically (black points in Figure 6d), reaching negative values above $I_{th}$. Moreover, the coercive field changes sharply above $I_{th}$ (green points in Figure 6f). In Figure 6c, we highlight the key observation that the hysteresis loop reverses sign above the threshold fluence, as indicated by the star symbols. This sign reversal is also clearly evident in the $\Delta\gamma^{dif}$ signals at 16 mT (Figure 6e), which represent the difference in MO signal dynamics at the positions marked by solid and open stars in Figure 6c. As shown in Figure 6f, $\Delta\gamma^{dif}$ signals change sign around $I_{th}$. Despite this sign change, the recovery of the $\Delta\gamma^{dif}$ signals remains unchanged and closely follows the dynamics of $\Delta T/T$ (see Supplementary Figure S13). The magnetic origin of the measured MO signals[40] is confirmed by the identical dynamics of $\Delta\gamma$ and $\Delta\beta$, as discussed in detail in Supplementary Note 2.8 and shown in Supplementary Figure S10. The observed reversal of the hysteresis-loop sign for pump intensities above $I_{th}$ cannot be explained by a simple thermal effect, as the static MO signal does not exhibit a sign reversal with increasing temperature (Figure 2d). Similarly, it cannot be accounted for by a simple reorientation of the magnetic easy axis (see Supplementary Note 3). Instead, the observed sign reversal indicates an additional laser-induced magnetic response. We also note that the pump-induced magnetic change is fully recovered when the pump intensity is reduced below $I_{th}$, ruling out any permanent modification of the sample by the laser pulses.

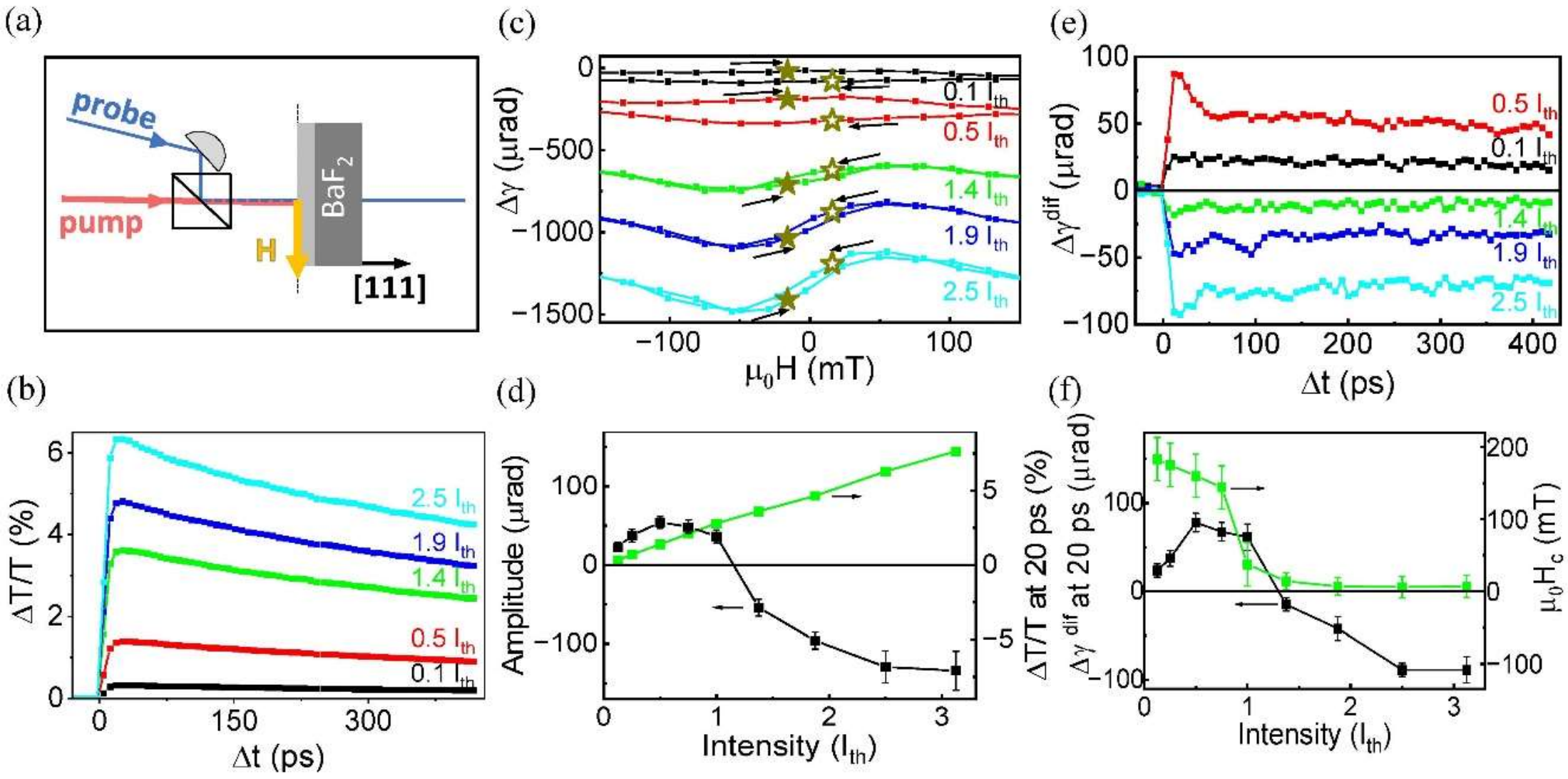

**Figure 6.** Intensity dependence measured using a concentric pump-probe setup, shown in **a**. **b**, Dynamics of transient transmission $\Delta T/T$. **c**, Pump-induced changes in hysteresis loops measured at $\Delta t$ = 20 ps; arrows indicate the direction of the magnetic field change (see also Supplementary Figure S12 where more data are shown). Solid (open) stars mark positions on the loop branch corresponding to measurements from negative (positive) fields, which were used to compute $\Delta\gamma^{dif}$ for 16 mT, as shown in e. **d**, Amplitude of hysteresis loops (see Supplementary Note 2.9 for the used sign convention) and $\Delta T/T$ signal at $\Delta t$ = 20 ps deduced from data depicted in c and b, respectively. **e**, Dynamics of $\Delta\gamma^{dif}$ for 16 mT. **f**, $\Delta\gamma^{dif}$ at $\Delta t$ = 20 ps for 16 mT and coercive field $H_c$ deduced from data depicted in e and c, respectively. Pump threshold fluence $I_{th} \approx$ 6.5 mJ cm$^{-2}$, wavelength 800 nm, circular polarization $\sigma$; probe wavelength 532 nm, polarization $\beta$ = 0°.

When the applied field exceeds the equilibrium coercive field, the situation changes significantly, as shown in **Figure 7** with data measured at 500 mT. Interestingly, unlike in the previously used geometries (Figure 3, 4, and 5), the $\Delta\gamma^{dif}$ signal is positive for all time delays (Figure 7a). The disappearance of the negative MO signal with a sub-picosecond rise time in this experimental geometry, where the longitudinal Kerr-like signal does not contribute to the measured MO signal, shows that the negative signal in the non-concentric setup is due to the ultrafast in-plane motion of magnetization, caused by spin-orbit torque. The precession amplitude initially increases with pump intensity but begins to decrease near $I_{th}$ (Figure 7c). Notably, the initial oscillation phase, which provides information about the mechanism triggering the magnetization precession,[36] does not depend significantly on intensity (Figure 7d), whereas the precession frequency, which reflects the sample's magnetic anisotropy, decreases with pump intensity. When the oscillatory signal is removed from the measured data by fitting with Equation 3, the remaining background signal (Figure 7b) corresponds to the out-of-plane position of the quasi-equilibrium easy axis about which the magnetization precesses in the laser-excited sample. In the concentric experimental setup, the detected MO signal corresponds solely to the MO Faraday effect, which is sensitive to the out-of-plane component of magnetization. Therefore, considering the static value of the MO coefficient (Figure 5f), the maximal MO signal of ≈ 100 µrad corresponds to an out-of-plane tilt of the easy axis of ≈ 5°.

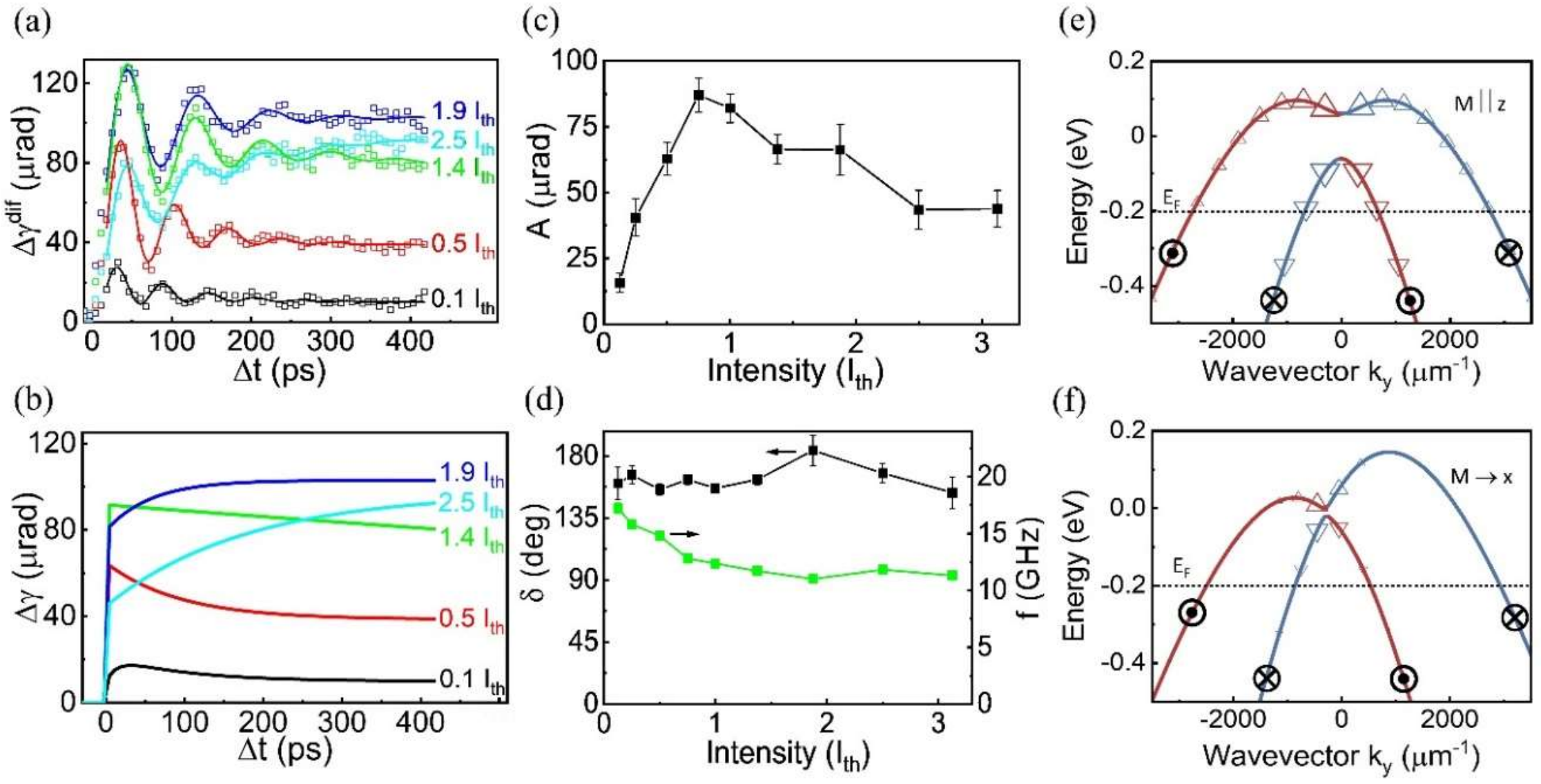

**Figure 7.** Intensity dependence measured using a concentric pump-probe setup for 500 mT. **a**, Dynamics of $\Delta\gamma^{dif}$ (points). The solid lines are fits using Equation 3 with precessional amplitude $A$, frequency $f$, and initial phase $\delta$ shown in **c**, and **d**; the precession damping time $\tau_1$ ≈ 70 ps does not depend on the intensity. **b,** Background signals, obtained by numerically removing the oscillations from the $\Delta\gamma^{dif}$ signals. Pump threshold fluence $I_{th}$ ≈ 6.5 mJ cm$^{-2}$, wavelength 800 nm, circular polarization $\sigma$; probe wavelength 532 nm, polarization $\beta$ = 0°. **e,** Computed band structure for magnetization in the out-of-plane $z$-direction, where it points without applied field. The black symbols depict the $y$-direction spin orientation. The triangles show the $z$-direction spin projection, with the spin orientation and projection magnitude encoded in the triangle orientation and size, respectively. The horizontal line represents the assumed position of the Fermi level. **f,** Same as e, for magnetization tilted by the applied field by 80° towards the $x$-direction.

### 2.4. Discussion

The magneto-optical data measured by pump-probe experiments in the GeMnTe sample are rather complex. However, evaluating the $\Delta\gamma^{dif}$ signal from the raw data makes it possible to separate genuine laser-induced magnetization dynamics from nonmagnetic optical background contributions. In this chapter, we present our interpretation of the results and place them in the broader context of the ultrafast manipulation of magnetic order in solids. A comparison of the results obtained for the GeMnTe sample with those reported for (Ga,Mn)As, the most thoroughly studied diluted magnetic semiconductor,[42] shows that they cannot be explained by a pump-induced temperature increase[31] or by the generation of strain pulses[44]. Instead, they are consistent with a transient increase in hole concentration that generates spin-orbit torque (see the detailed discussion in Supplementary Note 3). Moreover, for excitation fluences above the threshold fluence $I_{th} \approx 6.5$ mJ cm$^{-2}$, they are consistent with the reported pump-induced change in the sublattice shift in (Ge,Mn)Te,[14,16,17] which leads to a modification of the Rashba parameter.[14]

As background for these arguments, we show in Figure 7e and 7f the calculated band structures of (Ge,Mn)Te without and with an applied magnetic field, respectively (see Methods for a description of the theoretical model). Without an applied magnetic field (Figure 7e), the magnetization points along the inversion symmetry-breaking *z*-direction, opening a Zeeman gap at the degeneracy point through the exchange coupling of the holes with the Mn spin moments.[12,15,25] When the applied in-plane magnetic field tilts the magnetization toward the *x*-direction, the symmetry of the system changes, resulting in the band structure shown in Figure 7f, which explains why experimental data measured for 16 mT (Figure 6e) and 500 mT (Figure 7a show different intensity dependences.

There are two effects through which an intense laser pulse can influence the magnetization in the sample. First, absorption of the laser pulse generates additional holes in the system, leading to a transient shift of the Fermi energy. This changes the population of the spin-locked bands and, consequently, alters the magnetization easy axis. As a result, the magnetization begins to follow the easy axis shift through precessional motion. The intensity-independent initial phase observed for 500 mT (Figure 7d) shows that this hole-concentration-related excitation mechanism triggers the magnetization precession by a sub-picosecond in-plane tilt of the magnetization (inset in Figure 5c) for all intensities in our experiment. This mechanism is fully analogous to the optical spin-orbit torque observed in the diluted magnetic semiconductor (Ga,Mn)As[35] (see Supplementary Note 3 for a more detailed description of this effect). Moreover, the initial enhancement of the MO hysteresis amplitude with increasing laser intensity, which is apparent up to ≈ 0.5 $I_{th}$ (Figure 6d), can also be attributed to the increase in hole concentration, which enhances the hole-mediated magnetism in the GeMnTe sample, as observed in (Ga,Mn)As.[41,42]

The second mechanism by which the laser pulse can influence magnetic ordering in the GeMnTe sample is the laser-induced modification of the sublattice shift, as shown in Figure 1c.[14,16,17] The resulting change in the Rashba parameter leads to a change in the momentum offset between the bands with opposite spin-locked states, depicted in red and blue in Figure 7e and 7f. The modification of the band structure influences the magnetization, which can again be described as an effective spin-orbit torque. This scenario is consistent with the pronounced changes in the hysteresis loops observed in our experiment. The transition in the character of the hysteresis loops exhibits a clear threshold laser fluence, above which the loop amplitude increases sharply, accompanied by a reversal of its sign and a significant decrease of the coercive field. This effect saturates at fluences exceeding approximately 2.5 $I_{\mathrm{th}} \approx 16$ mJ cm$^{-2}$ (Figure S6b and Supplementary Figure S11b), of which approximately 65% is absorbed by the sample (see Methods). The threshold behavior and subsequent saturation of the measured magnetic response can be compared with laser-induced modifications of the GeTe sublattice displacement, where two distinct mechanisms have been reported in the literature. A surface-related enhancement of Rashba coupling was observed by time- and angle-resolved photoemission spectroscopy,[14] while a bulk-related suppression of the sublattice displacement was observed by ultrafast electron diffraction[16] and ultrafast X-ray diffraction[17] experiments. In our experiment, it is difficult to distinguish between these two mechanisms, as both modify the Rashba parameter and, consequently, the electronic band structure relative to equilibrium. However, the saturation fluence of the magnetic response in our experiments appears to be more consistent with the value reported for the bulk suppression of the Rashba parameter (Figure 3 in Ref. [16]) than with the substantially lower saturation fluence associated with the surface-related enhancement (Supplementary Figure 10 in Ref. [14]). Although a quantitative determination of the transient Rashba parameter is beyond the scope of the present study, the similar fluence dependence of our magnetic response and the previously reported laser-induced lattice displacement provides further support for our interpretation of the experimental data.

Ultrafast control of magnetic order using femtosecond laser pulses has attracted considerable attention since its discovery in 1996, when demagnetization of a thin Ni film in less than 1 ps was demonstrated.[18] These experiments enable the study of magnetic system dynamics on a time scale corresponding to the exchange interaction (tens to hundreds of femtoseconds), which is much shorter than the time scales of spin-orbit interaction (picoseconds) and magnetization precession (hundreds to thousands of picoseconds in ferromagnets). The laser-induced spin-orbit torques reported in this article are phenomenologically part of a class of nonthermal photomagnetic phenomena in which absorption of pump photons directly influences magnetic parameters, such as magnetocrystalline anisotropy.[18] In general, ultrafast excitation by laser pulses can provide access to new topological phases (e.g., skyrmions) that otherwise remain hidden in magnetic materials during adiabatic field cycling.[43] The substantial laser-induced change of the coercive field observed in the GeMnTe sample for intensities exceeding the threshold fluence (Figure 6f) appears to be another example of such phenomena. Remarkably, in the nanosecond time range, when the sublattice shift is fully recovered[14,16,17] and the perturbed properties of the sample are returning to equilibrium with a lifetime of $\approx$ 1 ns (Figure 5b), the magnetically modified material does not immediately return to the original equilibrium structure. This is evidenced by the time-delay-independent value of the coercive field in the nanosecond time range (inset in Figure 5d). Nevertheless, because the measured MO spectra are virtually identical in the equilibrium and laser-perturbed states (Figure 5f), they do not correspond to different magnetic phases. Instead, the nonequilibrium state appears to represent a reconfiguration of the correlated spin-glass structure of the GeMnTe sample into a lower-energy state.

## 3. Conclusion

In conclusion, we investigated the ultrafast magnetization dynamics of the multiferroic semiconductor $Ge_{0.85}Mn_{0.15}Te$ using time-resolved magneto-optical spectroscopy. By isolating the magnetic contribution to the transient optical response, we identified ultrafast changes in the magnetic state of our sample that can be interpreted as two distinct laser-induced magnetic phenomena: coherent magnetization precession and transient modifications of magnetic properties, evidenced by changes in both the magnitude and coercive field of the hysteresis loops.

The magnetization precession is triggered by a photoinduced increase in hole concentration, which alters the population of Rashba-split, spin-locked valence band states and consequently shifts the position of the magnetic easy axis. In contrast, the transient change in magnetic properties is attributed to laser-induced modifications of the sublattice displacement, which affect the ferroelectric polarization and the associated Rashba spin-orbit interaction. While the first type of optical spin-orbit torque has already been observed in the diluted magnetic semiconductor (Ga,Mn)As, the second effect had not been observed prior to this experiment in the multiferroic Rashba semiconductor (Ge,Mn)Te, highlighting the new physics arising from multiferroic coupling.

Our findings reveal two complementary pathways by which femtosecond optical excitation couples to the magnetic state of GeMnTe through its Rashba-coupled electronic structure. More broadly, they demonstrate that the strong interplay among charge carriers, ferroelectric polarization, spin-orbit coupling, and magnetization enables ultrafast control of magnetic properties in multiferroic semiconductors.

## 4. Methods

*Samples*: Two 50-nm-thick (Ge,Mn)Te layers with different Mn concentrations, $x_{Mn} = 0.15$ and 0.30, were grown by molecular beam epitaxy (MBE) on a $BaF_2$ substrate at 280°C, using GeTe, Mn, and Te beam flux sources. To protect the samples from oxidation, a 10 nm $Al_2O_3$ capping layer was deposited ex situ by atomic layer deposition.

The samples were structurally characterized by X-ray diffraction angular scans, which revealed a clear (Ge,Mn)Te peak corresponding to the rhombohedral structure and confirmed the high quality of the deposited layers. Reciprocal space maps (RSMs) measured around the $BaF_2$ (222) reflection at room temperature revealed that for the sample with $x_{Mn} = 0.15$ the (Ge,Mn)Te peak splits into at least two additional peaks due to the presence of ferroelectric domains oriented not only along the out-of-plane [111] direction, but also along the cubic body diagonals ⟨111⟩.[21] In contrast, no such splitting is observed for the sample with $x_{Mn} = 0.30$, since this (Ge,Mn)Te layer is not ferroelectric at room temperature. The magnetic properties of the sample with $x_{Mn} = 0.30$ were studied using a superconducting quantum interference device (SQUID) magnetometer. The temperature dependence of the magnetic moment revealed that the sample is magnetically ordered for temperatures up to ≈ 180 K, again confirming the high quality of this sample.

Further details about the sample growth and characterization are provided in Supplementary Note 1. and in Refs. [19-21].

*Pump-probe experiment*: Time-resolved pump–probe measurements were performed using two optical parametric amplifiers (Orpheus NEO, Light Conversion) pumped by a 20 W ytterbium-doped femtosecond laser (Pharos, Light Conversion), providing independent spectral tunability for the pump and probe pulses. The pulses had a duration of approximately 150 fs and could be tuned over a spectral range of 350–1100 nm at a repetition rate of 100 kHz. The sample was mounted on a cold finger of a closed-loop helium cryostat (ARS), with the temperature adjustable from 15 to 800 K.

Two types of pump-probe experiments were conducted.[23] In the "standard" geometry (Figure 1a), the pump and probe beams propagated non-collinearly. The pump beam was perpendicular to the direction of the applied magnetic field generated by an electromagnet, while the probe beam propagated at an angle of approximately 10 degrees relative to the pump beam. As shown in the inset of Figure 1a, the sample holder allowed tilting of the sample by an angle about the axis perpendicular to both the applied magnetic field and the propagation directions of the beams, resulting in different angles of incidence (AOI) of the probe beam; see Supplementary Figure S7 for a depiction of the experimental arrangement for several AOI values. The pump pulses were either linearly polarized, with the orientation of the polarization plane controlled by a half-wave plate and described by an angle $\alpha$ (see Figure 1a), or circularly polarized ($\sigma$), using a quarter-wave plate. The probe pulses were linearly polarized, with the orientation of the polarization plane controlled by a half-wave plate and described by an angle $\beta$. Both beams were focused onto the sample by independent lenses, resulting in beam diameters of approximately 40 μm and 20 μm (full width at half maximum, FWHM) for the pump and probe beams, respectively.

In a "concentric" geometry (Figure 6a), the pump and probe beams propagate in the same direction, perpendicular to both the applied magnetic field and the sample surface (AOI = 0°). This geometry is advantageous because it eliminates several contributions to the measured magneto-optical (MO) signals (see the main text), so only the MO Faraday effect contributes to the measured signal. However, using a dichroic beam splitter to combine both beams (see Figure 6a) prevents complete polarization control and broad spectral tunability of the pump and probe beams, as achieved in the "standard" setup. Both beams were focused by the same lens, resulting in a beam diameter of approximately 40 μm (FWHM) for each beam.

The pump-induced MO response change was detected by monitoring changes in probe ellipticity ($\Delta\gamma$) and rotation ($\Delta\beta$) as a function of the pump–probe time delay ($\Delta t$). The signal was measured using an optical bridge, where the MO contribution is obtained from the difference between the two photodetector outputs (see Appendix B in Ref. [45] for details). To characterize the dependence of the as-measured signals, for example $\Delta\gamma$, on the external magnetic field $H$, we computed signal components $\Delta\gamma^{dif}$ and $\Delta^{aver}$ that are defined as

$$\Delta\gamma^{dif} = [\Delta\gamma(-H) - \Delta\gamma(+H)]/2 \quad (1)$$

$$\Delta\gamma^{aver} = [\Delta\gamma(-H) + \Delta\gamma(+H)]/2\,. \quad (2)$$

Here, $\Delta\gamma^{dif}$ represents signals that change sign when the field direction is reversed, corresponding to odd-in-magnetization MO signals (such as the Faraday and Kerr effects). Similarly, $\Delta\gamma^{aver}$ represents signals that do not change sign when the field direction is reversed, that is signals not associated with magnetic ordering in the sample or corresponding to even-in-magnetization MO signals (such as the Voigt effect). Moreover, to account for magnetic hysteresis in the GeMnTe sample, we strictly distinguish between MO signals measured for different branches of the hysteresis loops. In this article, the notation $\Delta\gamma^{dif}$ and $\Delta\gamma^{aver}$ is used for signals computed from raw data for two different branches, always reaching the intended value of $H$ from the largest field magnitude of a given polarity (see the solid and open stars in Figure 6c). When MO signals measured within the same branch are used in Equation (1) or (2), the notation $\Delta\gamma^{dif*}$ and $\Delta\gamma^{aver*}$ is used instead (see Figure S5).

The oscillatory MO signals, which correspond to pump-induced precession of magnetization, were fitted by[36]

$$MO(\Delta t) = A\cos(2\pi f\Delta t + \delta)e^{(-\Delta t/\tau_1)} + B\left(1 - e^{(-\Delta t/\tau_2)}\right)e^{(-\Delta t/\tau_3)} + c\,, \quad (3)$$

where $A$, $f$, $\delta$, and $\tau_1$ are precessional amplitude, frequency, initial phase, and damping time, respectively. In addition, the background signal can described by $B$, $\tau_2$ and $\tau_3$ that represents the time-dependent background amplitude, rise time and decay time, respectively, and $c$ corresponds to the background that does not change in the measured time window.

Simultaneously with probe polarization changes, the sum of the detector signals in the optical bridge was recorded to monitor the pump-induced change in transmitted probe intensity. The relative transmission change was evaluated as $\Delta T/T = (T_e - T)/T$, where $T_e$ and $T$ denote the probe transmission with and without pump excitation, respectively.[45] The experiment was performed using pump pulses with a fluence ranging from approximately 0.8 to 27 mJ cm$^{-2}$; the fluence of the probe pulses was approximately 0.06 mJ cm$^{-2}$. The incident pump threshold fluence $I_{th}$ corresponds to approximately 6.5 mJ cm$^{-2}$, of which about 65% is absorbed in the $Ge_{0.85}Mn_{0.15}Te$ sample, as estimated from the measured transmission spectrum (Figure S1f) and the sample reflectivity. All experiments were performed at temperature 15 K.

*Static magneto-optical (MO) experiment*: Static characterization of the sample was performed using the experimental setup schematically shown in Figure 2b. A supercontinuum laser (SuperK EXTREME, NKT Photonics) was used as the light source, generating laser light across a broad spectral range (400–2400 nm). The desired wavelength was selected with a SuperK VARIA tunable filter. The sample was mounted on the cold finger of a closed-loop helium cryostat (ARS), allowing temperature variation from 15 to 800 K. The cryostat was placed in a custom-built four-pole electromagnet (see Ref. [34] for a detailed description of the experimental setup and the properties of the four-pole electromagnet).

In the implemented transmission experimental geometry, the laser beam was incident perpendicular to the sample plane (AOI = 0°), eliminating any potential contributions from MO effect analogous to the longitudinal MO Kerr effect.[39] The magnetic field was applied at different angles $\Theta_H$ relative to the sample plane (see Figure 2b). The signal was measured using an optical bridge, with the MO contribution obtained from the difference between the two photodetector outputs.[34] For selected values of $\Theta_H$, the sample's MO response was measured as a function of the applied magnetic field strength. In principle, the MO response of the cryostat windows, which are made from fused silica, can also contribute to the measured signals. To suppress this effect, we measured the sample-induced change in light ellipticity ($\gamma$) rather than polarization rotation ($\beta$), for which the MO response of fused silica is considerably stronger. Additionally, we used light wavelengths close to the spectral point where the fused silica ellipticity signal changes sign (approximately 700 nm), which further reduces the window contributions to the measured ellipticity signals. To suppress any potential contribution of the quadratic-in-magnetization MO Voigt effect[46] to the measured signals, we used light polarization oriented along the magnetic field projection onto the sample plane (i.e., vertical polarization in Figure 2b). Consequently, only the MO Faraday effect from the studied sample contributes to the measured signal in this experiment.

*Band structure calculations*: We considered non-relativistic kinetic energy where the band non-parabolicity is expressed in terms of the non-parabolicity parameter $\bar{\alpha} = 0.25$ within the Kane model

$$H = \frac{\sqrt{1+\frac{\bar{\alpha}\hbar^2 k^2}{m^*}}-1}{2\bar{\alpha}} + \frac{\Delta_Z}{2}\boldsymbol{m}\cdot\boldsymbol{\sigma} + \alpha_R\left(k_x\sigma_y - k_y\sigma_x\right), \quad (4)$$

where $k_j$ is the $j$-th component of the electron wave vector, $m^* = -0.32m_0$ is the effective electron mass with $m_0$ being the free electron mass, $\Delta_Z = 0.12$ eV is the Zeeman coupling parameter, $\alpha_R = 2 \times 10^{-4}$ eVµm is the Rashba parameter, $\sigma_j$ are the Pauli matrices and $\boldsymbol{m}$ is unit vector pointing along the magnetization. The energies and spin orientations of the eigenstates were obtained by diagonalization of the Hamiltonian above. The band structure shown in Figure 7e and 7f was calculated adapting the material parameters measured for a similar sample with $x = 0.13$.[12]

**Acknowledgements**

This work was supported by TERAFIT project No. CZ.02.01.01/00/22_008/0004594 funded by Ministry of Education Youth and Sports of the Czech Republic (MEYS CR), programme Johannes Amos Comenius (OP JAK), call Excellent Research, and by CzechNanoLab Research Infrastructure supported by MEYS CR (LM2023051). E.S. and H.R. acknowledge support by the Czech Science Foundation (Grant No. 22-17899K), DK acknowledges support by Lumina Quaeruntur fellowships LQ100102201 of the Czech Academy of Sciences, GS acknowledges support by Austria Science Funds, Project PIN6540324 and I 6568.

**Data Availability Statement**

Data reported in this paper are available in the Zenodo repository [10.5281/zenodo.22116843] and are publicly available as of the date of publication.

MO signals at these initial stages are dominated by "optical" rather than "magnetic" contributions, as discussed in Ref. [40] and Supplementary Note 2.8. This can be experimentally addressed by comparing the dynamics of pump-induced changes in ellipticity and rotation, as shown in Supplementary Figure S9 and S10. In GeMnTe, the strong probe-wavelength-dependent sensitivity of the measured signals to carrier relaxation within the band structure[7,11] makes probe wavelengths around 500 nm the best choice for studying early-stage magnetization dynamics, as shown in Supplementary Figure S9a.

**Supporting Information**

The online version contains supplementary material available at [to be provided]. Supplemental materials consist of Supplementary Note 1. Magnetic and structural characterization of studied samples (containing Figure S1), Supplementary Note 2. Additional experimental data (containing Figures S2-S13), and Supplementary Note 3. Comparison of results measured in (Ge,Mn)Te with results reported for (Ga,Mn)As.

# Supplementary information

## Laser-Induced Rashba Spin-Orbit Torques in Multiferroic Semiconductor (Ge,Mn)Te

*Zeynab Sadeghi, Tomáš Ostatnický, Eva Schmoranzerová, Jozef Kimák, Dominik Kriegner, Helena Reichlová, Lukáš Nádvorník, Gunther Springholz, J. Hugo Dil, Juraj Krempasky, and Petr Němec**

Contents

## Supplementary Note 1. Magnetic and structural characterization of studied samples

Two 50-nm-thick (Ge,Mn)Te layers with different Mn concentrations, $x_{Mn}$ = 15% and $x_{Mn}$ = 30%, were grown by molecular beam epitaxy (MBE) on $BaF_2$ substrate at 280°C, using GeTe, Mn, and Te beam fluxes [S1]. A protective $Al_2O_3$ capping layer was deposited ex situ.

### Structural characteristics

The samples were structurally characterized by X-ray diffraction angular scans (Fig. S1a). For both samples, a clear (Ge,Mn)Te peak corresponding to the rhombohedral structure is visible close to the $BaF_2$ (111) peak, confirming the high quality of the deposited layers. We further measured reciprocal space maps (RSMs) around the BaF2 (222) reflection at room temperature (Fig. S1b). For the sample with $x_{Mn}$ = 15%, the (Ge,Mn)Te peak splits into at least two additional peaks due to the presence of ferroelectric domains oriented not only along the out-of-plane [111] direction, but also along the cubic body diagonals ⟨111⟩ [S1]. In contrast, no such splitting is observed for the sample with $x_{Mn}$ = 30%, since this (Ge,Mn)Te layer is not ferroelectric at room temperature.

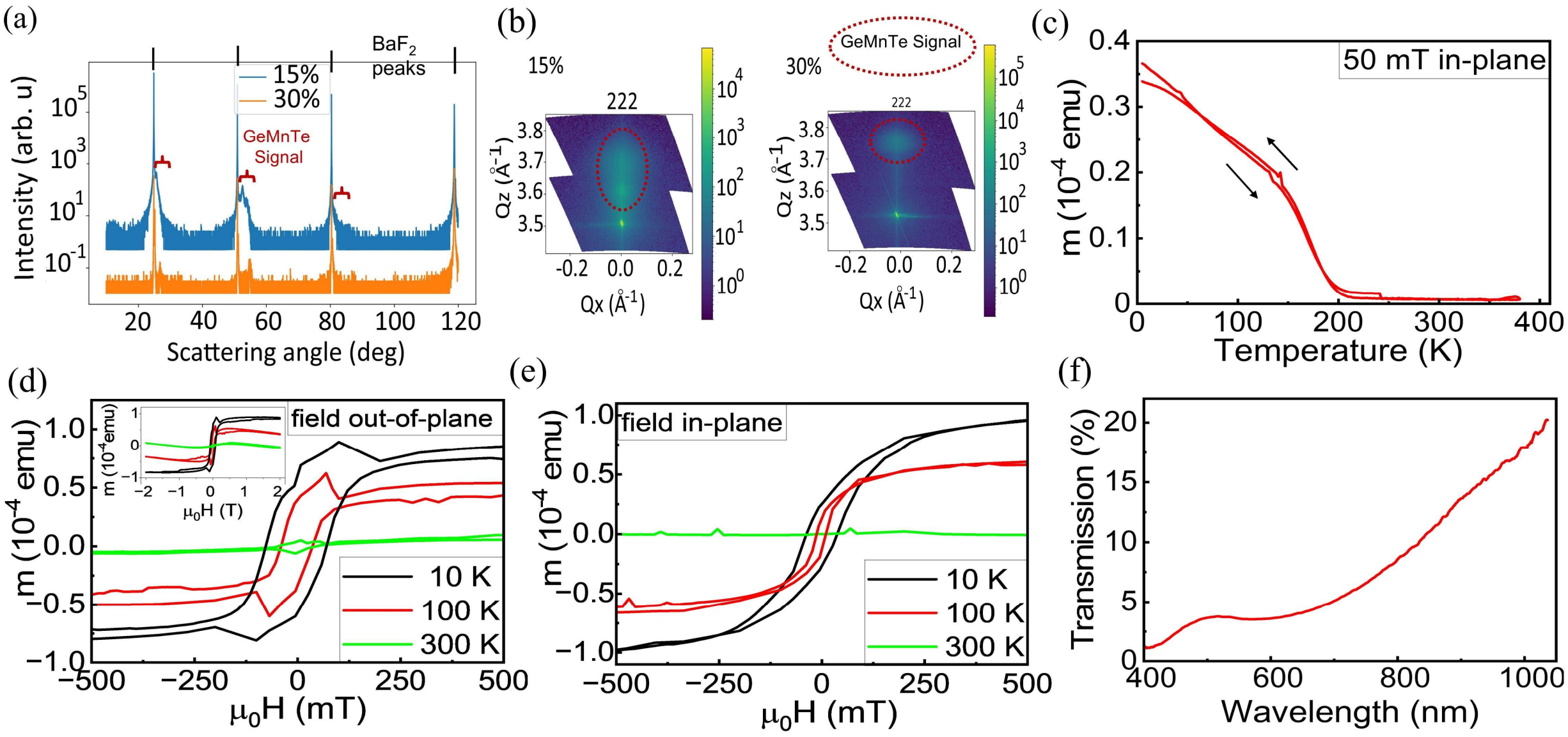

**Fig. S1: Characterization of (Ge,Mn)Te samples. a,** X-ray diffraction angular scans for samples with $x_{Mn}$ = 15% (blue curve) and 30% (orange curve). Positions of the (Ge,Mn)Te peaks are indicated. **b**, Reciprocal space maps around the (222) BaF2 direction $x_{Mn}$ = 15% (left panel) and 30% (right panel). Signal corresponding to (Ge,Mn)Te is indicated by red ellipse. **c**, Temperature dependence of magnetic moment *m*, measured with out-of-plane magnetic field of 50 mT, in the sample with 15 % of Mn. **d**, and **e**, Hysteresis loops in the out-of-plane (d) and in plane (e) directions. Inset in (d) shows the data in the full range of external magnetic fields. **f**, Transmission spectra of the samples with $x_{Mn}$ = 15% at 15 K.

### Magnetic characteristics

The magnetic properties of the 15% Mn sample were studied using a superconducting quantum interference device (SQUID) magnetometer. The temperature dependence of the magnetic moment, measured in an out-of-plane magnetic field of 50 mT, is shown in Fig. S1c.

The Curie temperature determined from this measurement is similar to the value obtained from magneto-optical characterization (Figure 2e). From the hysteresis loops recorded with the magnetic field applied in the out-of-plane (OOP) (Fig. S1d) and in-plane (IP) (Fig. S1e) directions, we conclude that the sample exhibits dominant OOP magnetic anisotropy. The inset of Fig. S1c shows that the OOP magnetization saturates only at relatively large magnetic fields, around 1.5 T at 10 K.

## Hole concentration

The hole concentration in our sample was estimated based on previous experiments reported in Ref. S1, Supplemental Fig. S3, where the hole concentration $p$ is plotted as a function of the Mn concentration $x_{\mathrm{Mn}}$. Those measurements were performed on nominally identical samples prepared in the same MBE chamber. We, therefore, estimate that for $x_{\mathrm{Mn}} = 15\%$, $p \approx 3 \times 10^{21}\mathrm{cm}^{-3}$.

## Transmission spectra

Figure S1f shows the transmission spectrum of the sample with $x_{Mn}$ = 15% at 15 K. This spectrum agrees with the spectral dependence expected for an indirect-bandgap semiconductor.

## Supplementary Note 2. Additional time-resolved magneto-optical response

This supplementary note contains additional time-resolved magneto-optical results supporting the interpretation provided in the main paper.

### 2.1 Polarization dependence

Figure S2 compares the magnetic-field-dependent ellipticity signals measured for different polarizations of pump and probe pulses. The signal obtained for circularly polarized excitation closely follows the average response measured for the two orthogonal linear pump polarizations $\alpha = 45°$ and 135°, as shown in Fig. S2a. The difference between these two orthogonal linear pump polarizations is nearly field independent and very small compared with the main magneto-optical response, see Fig. S2b. This indicates that, within the sensitivity of this measurement, the field-dependent ellipticity signal is not strongly affected by the pump polarization.

A similar comparison was performed for the probe polarization dependence. The response measured for $\beta = 0°$ is almost identical to the average of the signals measured for $\beta = 45°$ and 135° (Fig. S2c), showing that the dominant field-dependent contribution is largely independent of the selected probe polarization. The existing difference between the data measured for $\beta = 45°$ and 135° (Fig. S2d) can be attributed to Voigt magneto-optical effect, which is quadratic-in-magnetization.

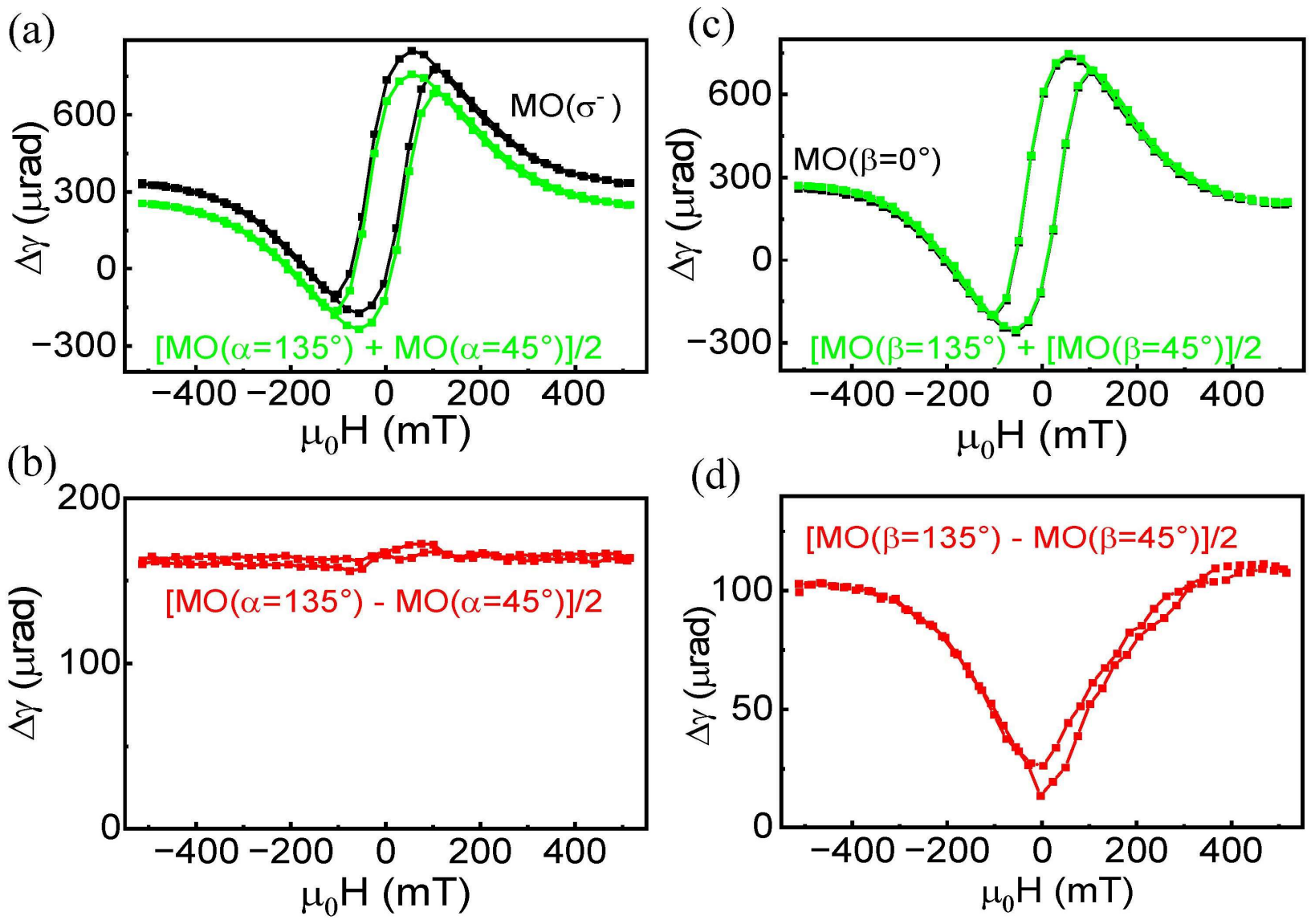

**Fig. S2: Pump- and probe-polarization dependences.** **a**, Field-dependent ellipticity $\Delta\gamma$ measured for $\Delta t = 20$ ps for circular pump polarization $\boldsymbol{\sigma}^{-}$ is compared with the average of signals measured for linear pump polarizations $\alpha = 45°$ and 135°. **b**, Difference between the signals measured for pump polarizations $\alpha = 45°$ and 135°. **c**, Signal measured for probe polarization $\beta = 0°$ is compared with the average of the signals measured for probe polarizations $\beta = 45°$ and 135°. **d**, Difference between the signals measured for probe polarizations $\beta = 45°$ and 135°. Measurements were performed in the standard geometry for pump fluence 1.3 $I_{th}$, wavelength 820 nm, probe wavelength 633 nm, AOI = -5°, $\Theta_H = -15°$.

## 2.2 Experiment below and above magnetic ordering temperature

At 15 K, the magnetic-field-dependent ellipticity loops show a clear hysteretic response for all pump polarizations, see Fig. S3a. The overall signal contains a polarization-dependent field-independent offset: the largest positive background is observed for $\alpha$ = 135°, while the largest negative background is observed for $\alpha$ = 45°. As discussed in the Methods section of the main text, we separate the dynamic into two components, $\Delta\gamma^{\mathrm{dif}}$ and $\Delta\gamma^{\mathrm{aver}}$. The differential part $\Delta\gamma^{\mathrm{dif}}$ contains the part of the signal that changes sign when the magnetic field is reversed. In contrast, the $\Delta\gamma^{\mathrm{aver}}$ component contains the part of the signal that does not change the sign under the field reversal.

Using this separation procedure, the $\Delta\gamma^{\mathrm{dif}}$ ellipticity dynamics measured at 500 mT are nearly identical for all pump polarizations, indicating that the magnetic part of the transient MO response does not depend strongly on the pump polarization (Fig. S3b). In contrast, the $\Delta\gamma^{\mathrm{aver}}$ component shows a pronounced pump-polarization dependence, following the same trend as the offset observed in the hysteresis loops (Fig. S3c).

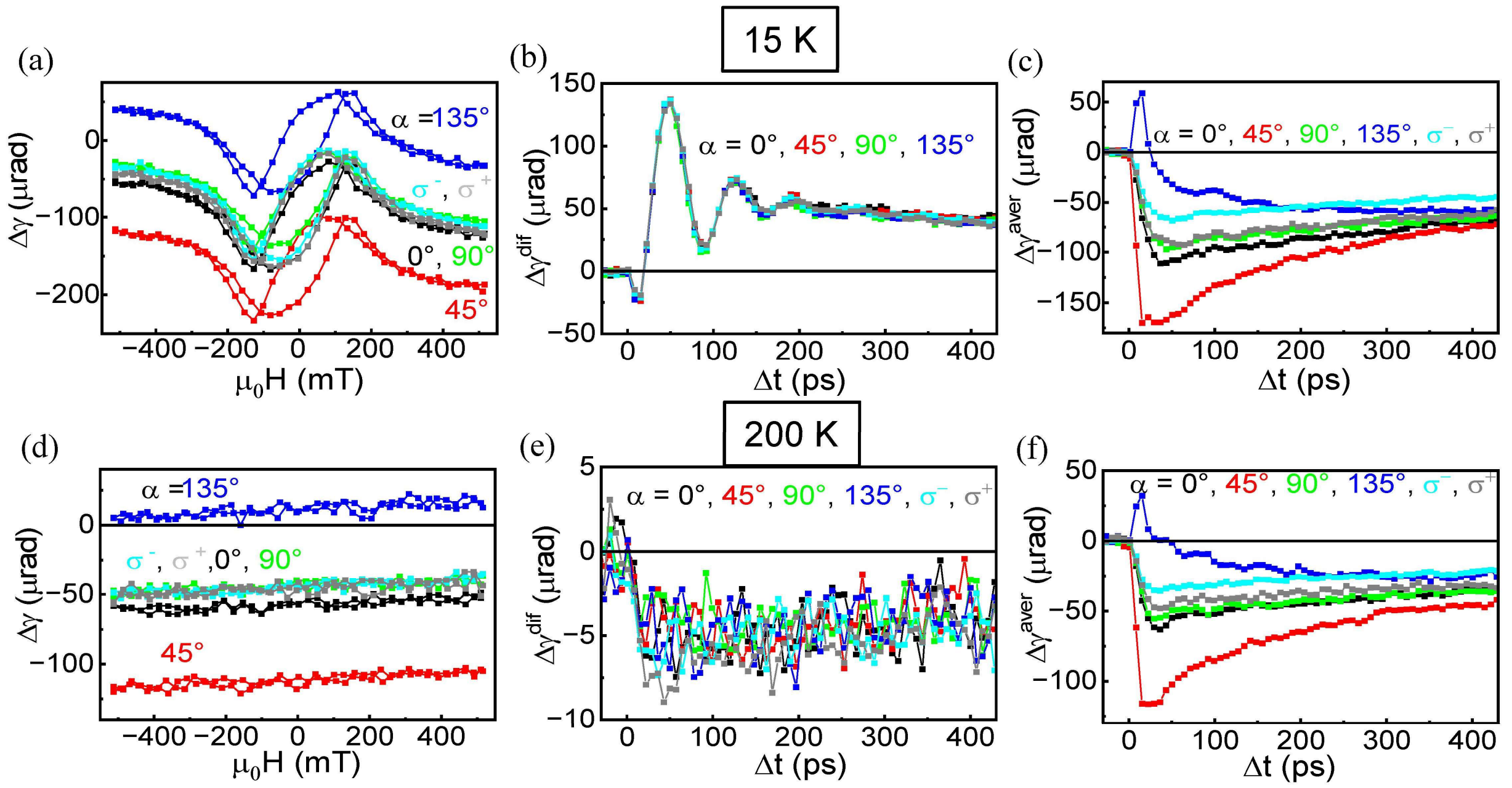

**Fig. S3: Pump-polarization dependence of MO signals at 15 K and 200 K. a**, Magnetic-field-dependent ellipticity hysteresis loops measured at 15 K for $\Delta t$ = 20 ps for different linear pump polarizations $\alpha$ = 0°, 45°, 90°, and 135°, and for circularly polarized excitations σ− and σ+.
**b**, and **c**, Dynamics of signals $\Delta\gamma^{dif}$ (b) and $\Delta\gamma^{aver}$ (c), see Methods for their definitions, measured at 15 K for 500 mT as a function of time delay for different pump polarizations. **d**, **e**, and **f**, Same as a, b, and c, respectively, measured at 200 K. All measurements were performed in the standard geometry for pump wavelength 820 nm, probe wavelength 633 nm, AOI = 0°, $\Theta_H$ = -10°.

The measurements at 200 K, which is above the magnetic transition temperature in the studied sample, provide a direct evidence how the measured signals change due to the quenched magnetic ordering in the sample. At this temperature, the hysteretic field-dependent signal is completely absent (Fig. S3d), and the component $\Delta\gamma^{dif}$ is reduced to a very weak signal (Fig. S3e). This confirms that the $\Delta\gamma^{dif}$ signal is connected with the magnetic order. On the other hand, the component $\Delta\gamma^{aver}$ is not affected significantly by the temperature increase and still shows pump-polarization-dependent signals (Fig. S3f). This suggests that the $\Delta\gamma^{aver}$ signal is mainly related to a pump-induced non-magnetic optical background rather than to the magnetic signal detected through the field-even MO Voigt effect.

### 2.3 Experiment in $Ge_{0.70}Mn_{0.30}Te$ reference sample

Reference time-resolved pump–probe experiments were performed in the $Ge_{0.70}Mn_{0.30}Te$ film. As shown in Figure S4a, the magnetic-field-dependent ellipticity measured for different pump polarizations show strong suppression of the hysteretic response. Similarly, the pump-polarization-independent $\Delta\gamma^{dif}$ signal (Fig. S4b) is rather small in this sample. On the other hand, there still a rather pronounced pump-polarization-dependent offset in the hysteresis loops (Fig. S4b) together with a rather large pump-polarization-dependent $\Delta\gamma^{aver}$ signal (Fig. S4c).

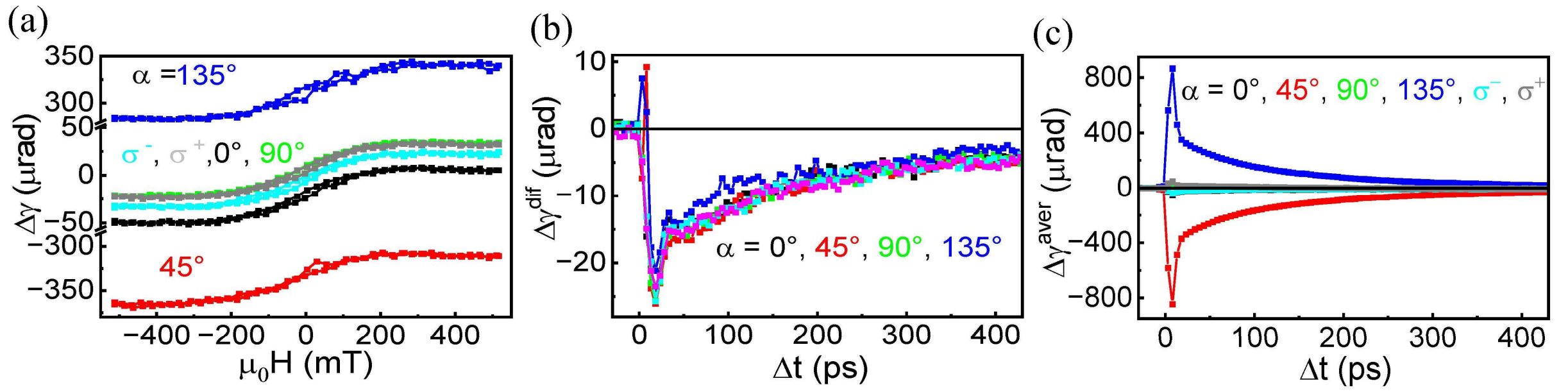

**Fig. S4: Pump-polarization dependence of MO signals measured at 15 K in $Ge_{0.70}Mn_{0.30}Te$.** **a**, Magnetic-field-dependent ellipticity hysteresis loops measured for $\Delta t = 20$ ps for different linear pump polarizations $\alpha = 0°$, 45°, 90°, and 135°, and for circularly polarized excitations σ− and σ+. **b**, and **c**, Dynamics of signals $\Delta\gamma^{dif}$ (b) and $\Delta\gamma^{aver}$ (c) measured for 500 mT as a function of time delay for different pump polarizations. All measurements were performed in the standard geometry for pump wavelength 820 nm, probe wavelength 633 nm, AOI = 0°, $\Theta_H = -10°$.

### 2.4 Time-resolved MO data within one hysteresis loop branch

Figures S5b and S3c show evolution of time-resolved signals when the magnetic field is swept from positive to negative fields within one hysteresis loop branch (see Fig. S5a).

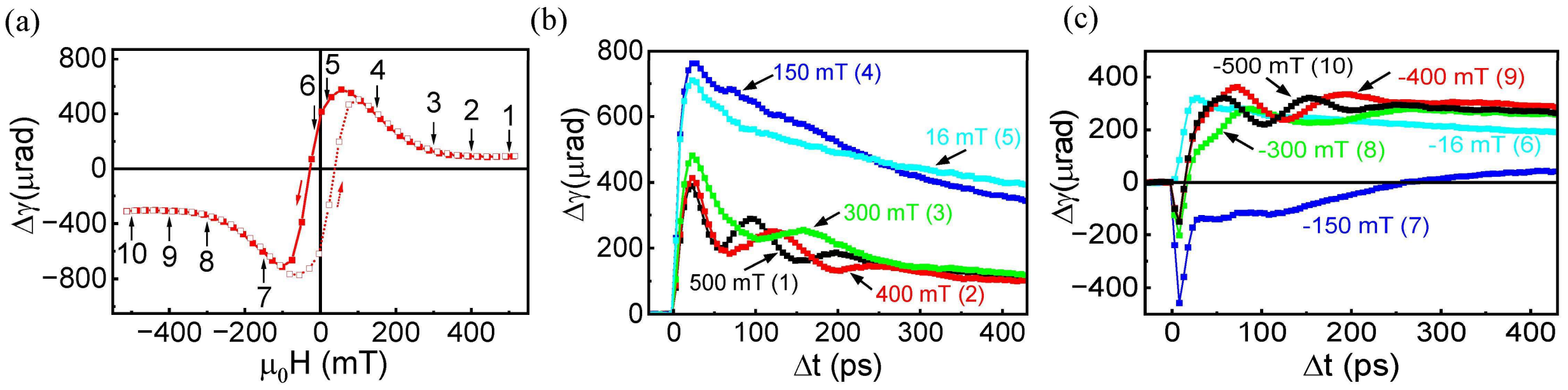

**Fig. S5: Magnetic-field dependence of the time-resolved MO signal.** **a**, Magnetic-field-dependent ellipticity hysteresis loops measured for $\Delta t = 2$ ps; the numbers by the vertical arrows indicate the measurements sequence, starting from +500 mT (1) and ending at −500 mT (10).
**b**, and **c**, Dynamics of $\Delta\gamma$ measured at fields depicted in a. All measurements were performed in the standard geometry for pump fluence 1.3 $I_{th}$, wavelength 820 nm, probe wavelength 633 nm, AOI = -5°, $\Theta_H = -15°$.

## 2.5 Precession of magnetization

Figure S6 shows the magnetic-field dependence of $\Delta\gamma^{dif*}$ signal, corresponding to a difference of MO signals measured at depicted fields within one branch of the hysteresis loop (see Figure S5a and Methods). The squares are the experimental data and the red solid lines are fits. The dynamics were fitted using equation 3 described in the Methods section where the first term describes the damped oscillatory response and the second term accounts for the non-oscillatory background. The fits reproduce the experimental data well for all three magnetic fields. The oscillation period becomes longer as the magnetic field decreases, showing that the precession frequency is reduced at lower fields (Fig. S6d). The values of other fitting parameters are shown in Figs. S6e and S6f. The oscillation amplitude $A$ is nearly field-independent while the background amplitude $B$ decreases for higher fields. The precession phase $\delta$ changes only weakly with magnetic field, and the decay time $\tau_1$ shows a moderate decrease at higher fields.

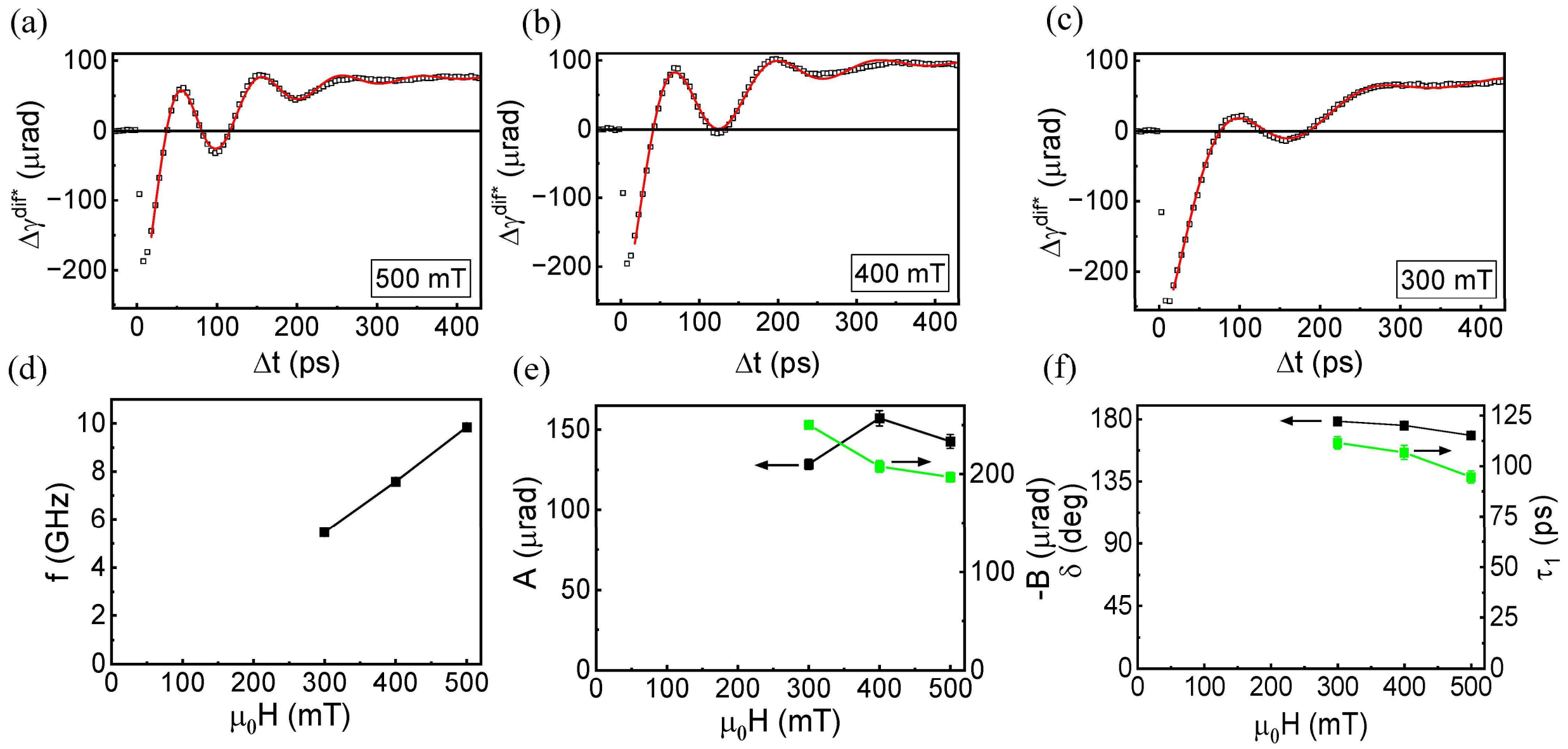

**Fig. S6: Time-resolved MO signals for selected magnetic fields. a**, **b**, and **c**, Dynamics of $\Delta\gamma^{dif*}$ measured for 500 mT, 400 mT, and 300 mT, respectively (points). Red solid lines show the fits by Equation 3 (see Methods). **d**, **e**, and **f**, Magnetic-field dependence of the fitted parameters describing the measured signals: frequency $f$ (d), oscillation amplitude $A$ together with the background amplitude $B$ (e), and phase $\delta$ together with the decay time $\tau_1$ (f). All measurements were performed in the standard geometry for pump fluence 1.3 $I_{th}$, wavelength 820 nm, probe wavelength 633 nm, AOI = -5°, $\Theta_H$ = -15°.

## 2.6 Angle-of-incidence dependence

Figure S7 shows how the obtained signals depend on the probe angle of incidence (AOI). In the experiment, the angle of incidence was varied by rotating the sample, which also changed the angle between the applied magnetic field and the sample plane (see inset in Fig. 1a). The three representative geometries are shown in Figs. S7a, S7b and S7c for AOI = −5°, 0° and +5°, respectively. The corresponding magnetic-field-dependent ellipticity loops are shown in Fig. S7d for several several values of AOI. Clearly, the shape and amplitude of the hysteresis loops change with AOI, indicating that the measured magneto-optical response is sensitive to the experimental geometry. The dynamics of $\Delta\gamma^{dif}$ and $\Delta\gamma^{aver}$ are shown in Fig. S7e and Fig. S7f, respectively. For all measured angles, the $\Delta\gamma^{dif}$ signal contains a damped oscillatory contribution, although its amplitude and frequency vary with AOI, see Figs. S7g, S7h and S7i. On the other hand, the $\Delta\gamma^{dif}$ signal does not depend on AOI significantly.

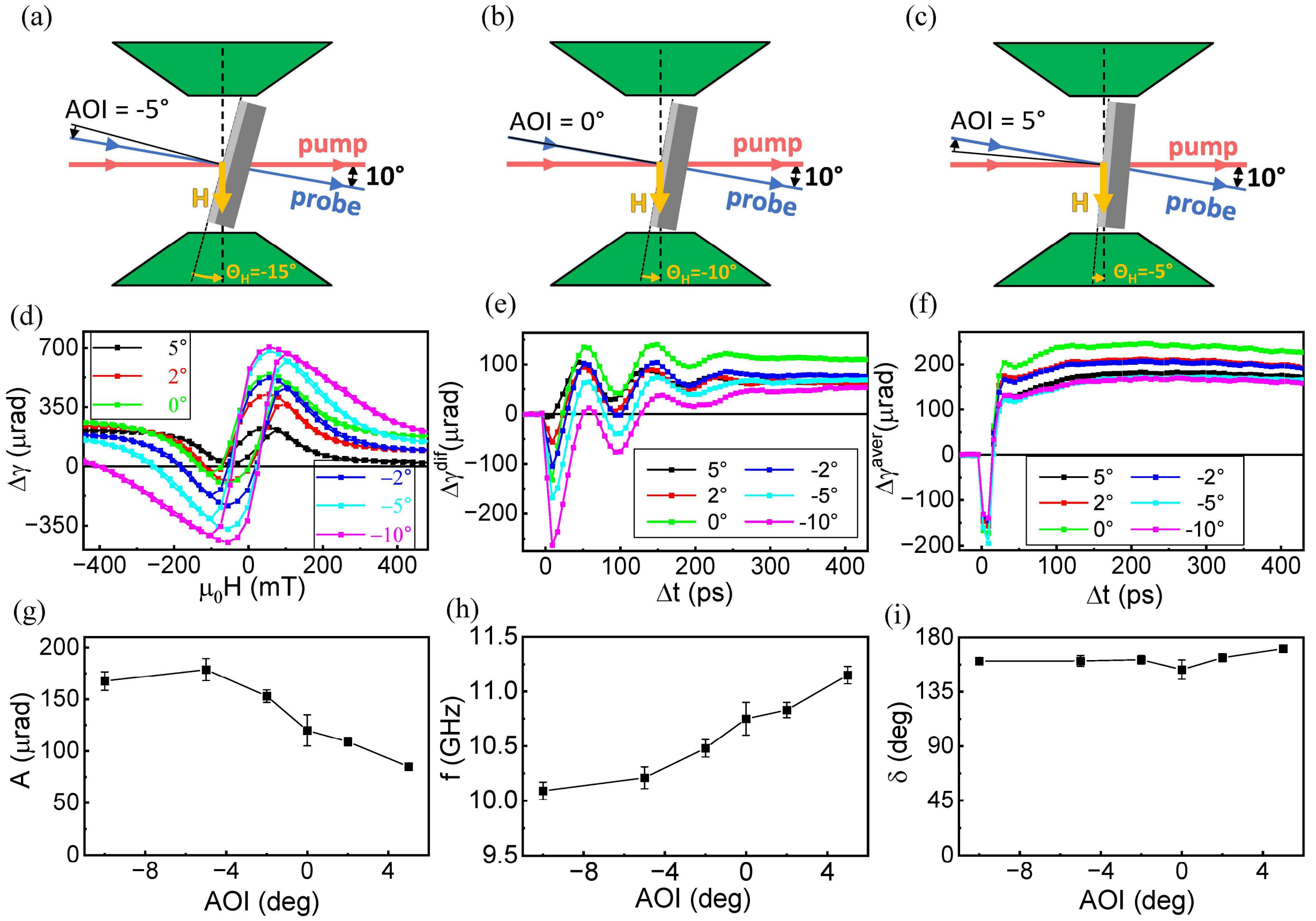

**Fig. S7: Influence of the angle of incidence (AOI) on the measured signals.**
**a**, **b**, and **c**, Schematic depiction of experimental geometries for different values of AOI and the resulting magnetic-field tilt angles $\theta_H$. a: AOI = −5°, $\theta_H$ = −15°, b: AOI = 0°, $\theta_H$= −10°, c: AOI = +5°, $\theta_H$ = −5°. **d**, Magnetic-field-dependent ellipticity hysteresis loops measured for $\Delta t$ = 20 ps at depicted values of AOI. **e**, and **f**, Dynamics of signals $\Delta\gamma^{dif}$ (e) and $\Delta\gamma^{aver}$ (f) obtained at depicted values of AOI. **g**, **h**, and **i**, Dependence of the fitting parameters, used to describe the magnetization precession in e, by Equation 3, on AOI: precession amplitude $A$ (g), frequency $f$ (h), and phase $\delta$ (i). All measurements were performed in the standard geometry for pump fluence 1.3 $I_{th}$, wavelength 820 nm, probe wavelength 633 nm.

## 2.7 Pump-wavelength dependence

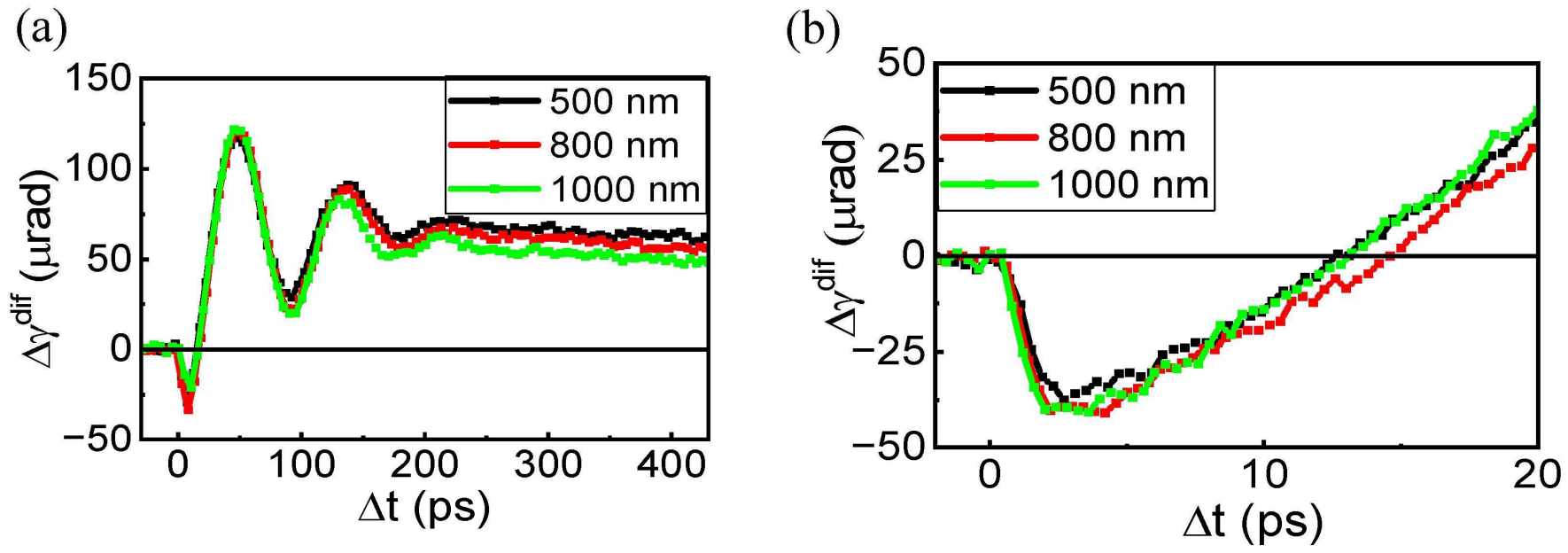

**Fig. S8: Pump-wavelength dependence of the $\Delta\gamma^{dif}$ signal**. **a**, and **b**, Dynamics of $\Delta\gamma^{dif}$ signal measured for pump wavelengths of 500, 800, and 1000 nm over the intermediate (a) and short (b) time ranges, respectively. All measurements were performed in the standard geometry probe wavelength 633 nm, AOI = 0°, applied magnetic field 500 mT. At each wavelength, the pump fluence was adjusted to obtain a fixed value of the transient transmission, which eliminates the role of spectrally dependent absorption coefficient in the sample.

Figure S8 shows the dependence of the $\Delta\gamma^{dif}$ signal dynamics on the pump wavelength. For all pump wavelengths, the dynamics show very similar behavior, including the initial rise, the damped oscillatory response, and the subsequent slow background evolution. This indicates that the magnetic dynamics are not strongly affected by changing the pump wavelength within the measured range.

## 2.8 Magnetic origin of MO signals

Magneto-optical (MO) investigation of the magnetization state and its dynamics relies on measuring changes in the polarization state of probe light – specifically, its rotation and change in ellipticity – induced by interaction with a sample. In principle, there are two contributions to the measured signals: one arises from symmetry breaking due to magnetization, and the other from non-magnetic optical anisotropy of the sample. In this article, we focus on the $\Delta\gamma^{dif}$ and $\Delta\beta^{dif}$ signals, which correspond to the signal components that change sign when the direction of the external magnetic field is reversed (see Eq. (1) in Methods). Therefore, this data processing completely removes the non-magnetic component from the raw data, leaving only the MO response of the sample.

Nevertheless, pump-induced MO signals are sensitive not only to the magnetic dynamics of the system under study but also to laser-induced changes in the complex refractive index. This contribution is often referred to as the optical part of the MO signal. To determine whether the magnetic or optical part dominates in the measured dynamical data, both the pump-induced changes in rotation $\Delta\beta$ and ellipticity $\Delta\gamma$ should be measured, and their dynamics compared [S2].

In the static case, when only the Faraday effect is present, rotation $\beta$ and ellipticity $\gamma$ could be described as

$$\beta = f_\beta M, \tag{S1}$$

$$\gamma = f_\gamma M, \tag{S2}$$

where $f_\beta$ and $f_\gamma$ are optical ”constants”, which dependent on material electronic properties and light wavelength, and $M$ is magnetization. If a pump pulse induces a material change in the studied sample, it leads to a time-dependent change of polarization rotation $\Delta\beta(t)$ and ellipticity $\Delta\gamma(t)$. Since both optical and magnetic parts could be modified by the pump pulse, we have

$$\Delta\beta(t) \approx \Delta f_\beta(t)M + f_\beta \Delta M(t)\ , \tag{S3}$$

$$\Delta\gamma(t) \approx \Delta f_\gamma(t)M + f_\gamma \Delta M(t)\ . \tag{S4}$$

The first terms on the right-hand sides are the optical parts, the second terms are the magnetic parts. One can see that in general case the dynamics of rotation and ellipticity does not need to be equal. However, if the magnetic parts dominate, i.e., if $\Delta f_\beta(t)M \ll f_\beta \Delta M(t)$ and $\Delta f_\gamma(t)M \ll f_\gamma \Delta M(t)$, then

$$\Delta\gamma(t) \approx \frac{f_\gamma}{f_\beta}\Delta\beta(t) \approx f_\gamma \Delta M(t)\ . \tag{S5}$$

This means that for $\Delta\beta(t)$ and $\Delta\gamma(t)$ to represent magnetic dynamics, the dynamics of rotation and ellipticity must be equal, up to a multiplicative constant. Similar analysis shows that measurements of ellipticity (or rotation) dynamics for various probe wavelengths must lead to the same dynamics, up to a multiplicative constant.

In Figure S9 the ellipticity $\Delta\gamma^{dif}$ and rotation $\Delta\beta^{dif}$ signals in the standard measurement geometry are compared. In Figure S9a and S9b the dynamics measured at probe wavelengths of 450 nm and 550 nm are shown, respectively. To enable a direct comparison of these dynamics, the rotation dynamics were multiplied by the depicted scaling factors. At both wavelengths, the rotation and ellipticity signals follow very similar time dependences, including the initial rise and the following relaxation. This shows that both these magneto-optical detection channels are sensitive to the same pump-induced change of magnetic properties in GeMnTe sample for time delays ≥ 1 ps, where the signals start to decay. This also reveals that the pump-induced change of magnetic properties is fully completed in less than 1 ps.

For comparison, the transient transmission $\Delta T/T$ measured at several probe wavelengths is shown in Fig. S9c. On time scales up to ≈ 10 ps, the signal strongly depends on the probe wavelength, in both sign and amplitude, reflecting the relaxation of photoexcited carriers within the band structure and their trapping at defect states. This leads to a time- and wavelength-dependent evolution of the complex refractive index of the excited film. This effect is most pronounced for time delays up to ≈ 1 ps, when the carriers lose their excess energy, which is why optical effects dominate in Equations (S3) and (S4) in this time range.

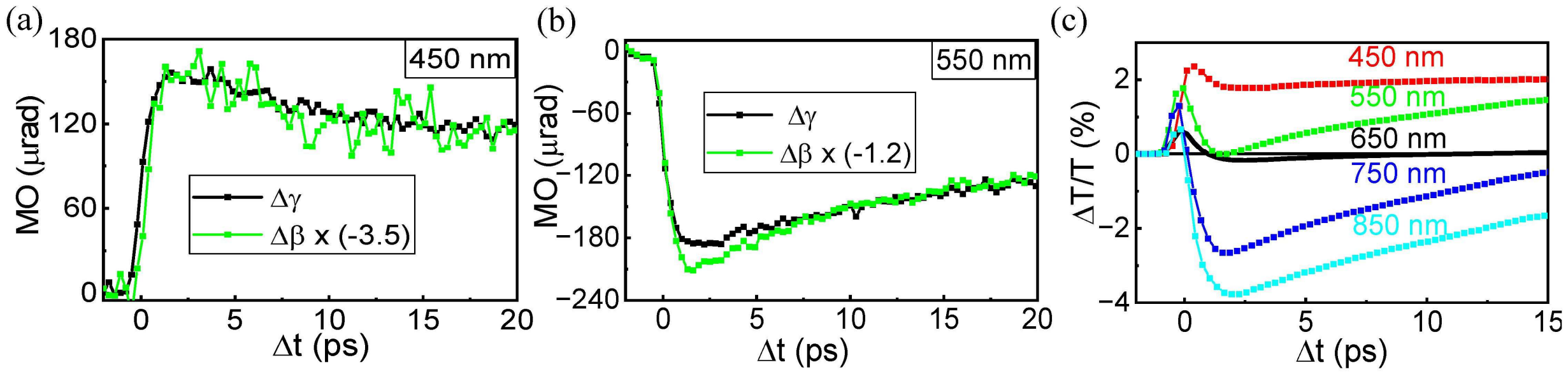

**Fig. S9: Comparison of rotation and ellipticity dynamics in the standard geometry.**
**a**, Comparison of the ellipticity $\Delta\gamma^{dif}$ and rotation $\Delta\beta^{dif}$ signals measured at a probe wavelength of 450 nm. The rotation signal is multiplied by a factor −3.5 to enable a direct comparison with the ellipticity data. **b**, Same as in a, for a probe wavelength of 550 nm; the rotation signal is multiplied by a factor −1.2. **c**, Transient transmission $\Delta T/T$ measured for different probe wavelength. Pump wavelength 1000 nm, AOI = 0°, applied magnetic field 16 mT.

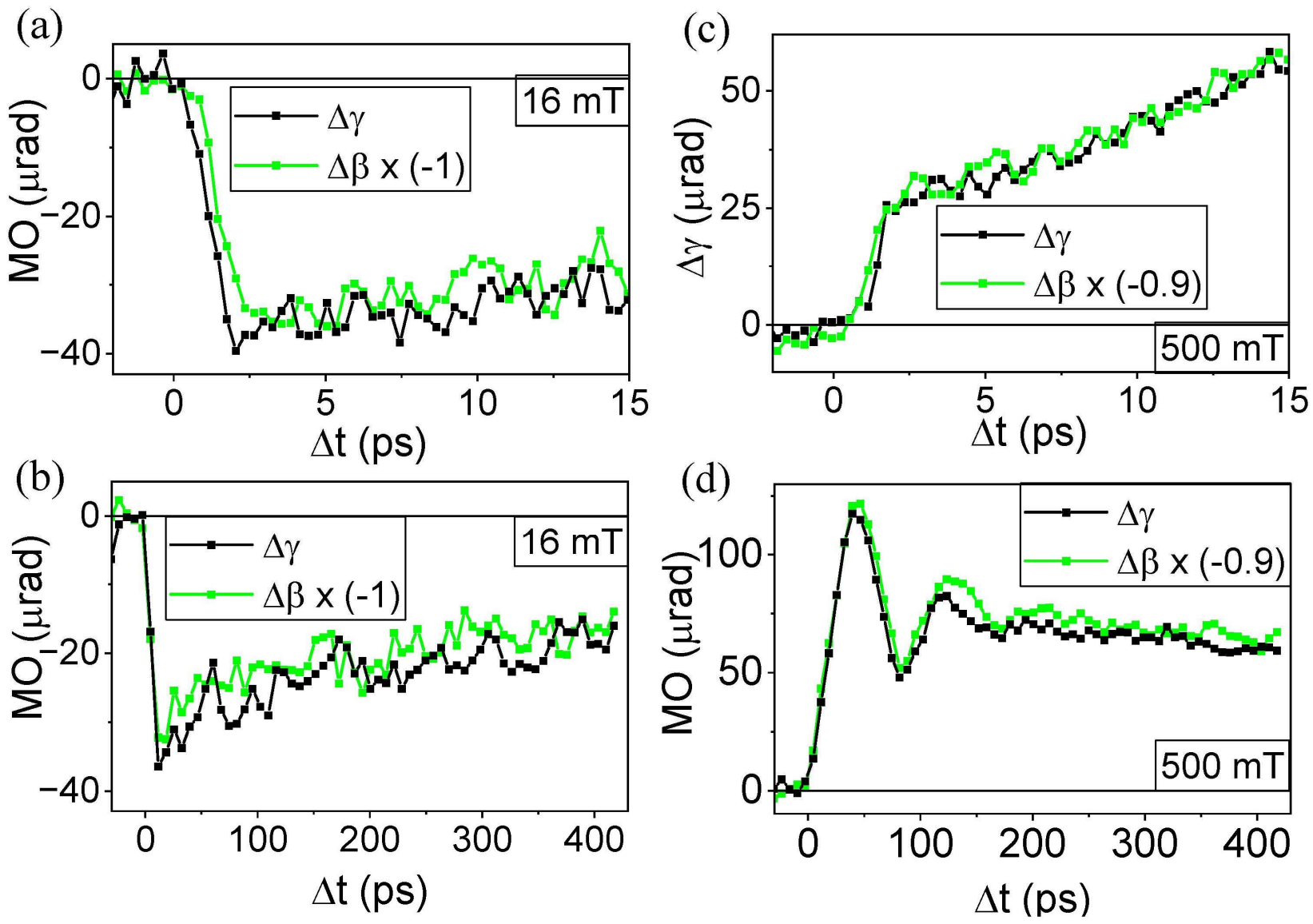

**Fig. S10: Comparison of rotation and ellipticity dynamics in the concentric geometry.**

**a**, and **b**, Comparison of the short (a) and intermediate (b) dynamics of ellipticity $\Delta\gamma^{dif}$ and rotation $\Delta\beta^{dif}$ signals measured at 16 mT. **c**, and **d**, Same as a, and b, for 500 mT. Pump wavelength 800 nm, probe wavelength 532 nm.

Figure S10 compares the ellipticity $\Delta\gamma^{dif}$ and rotation $\Delta\beta^{dif}$ signals in the concentric geometry at low and high magnetic fields. As for the standard geometry, this indicates that both magneto-optical detection channels are sensitive to the same pump-induced change of magnetic properties in GeMnTe sample for time delays ≥ 1 ps. Moreover, these results also verify that the pump-induced change of magnetic properties is fully completed in ≤ 1 ps.

## 2.9 Intensity dependence

Figure S11 shows how the magnetic-field-dependent ellipticity loops evolve with the pump intensity in the standard geometry. When the pump intensity is increased above the threshold fluence $I_{th}$, the loop amplitude grows rapidly, indicating a pronounced enhancement of the transient magneto-optical response. However, the measured signal approaches saturation at higher pump intensities.

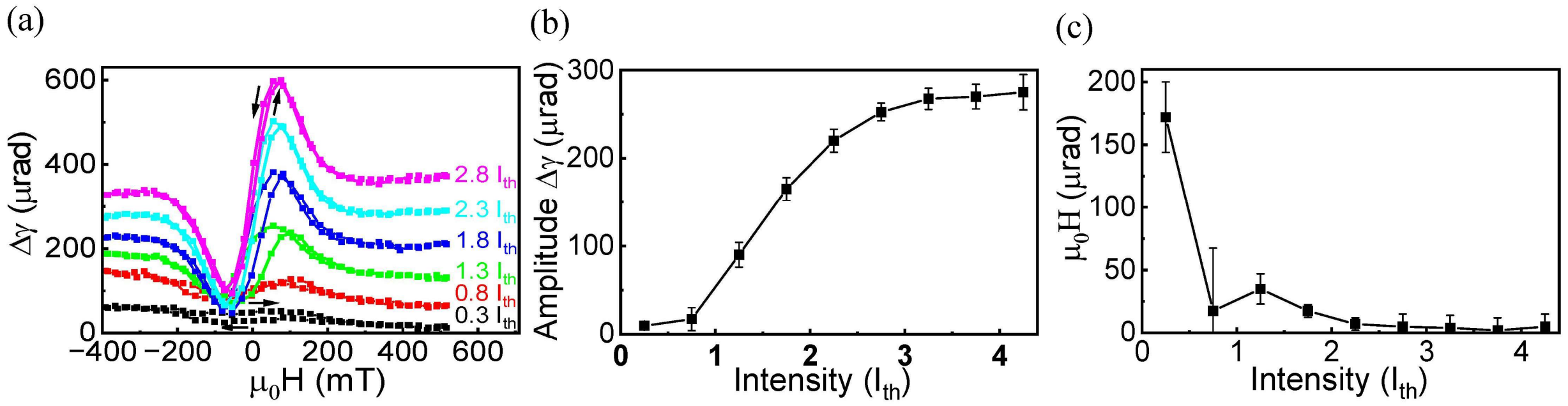

**Fig. S11: Pump-intensity dependence in the standard geometry.** **a**, Magnetic-field-dependent ellipticity hysteresis loops measured for $\Delta t$ = 20 ps at different pump intensities. **b**, and **c**, Pump-intensity dependence of the hysteresis-loop amplitude (b) and the coercive field (c) extracted from data shown in panel a. , Pump wavelength 1 000 nm, probe wavelength 670 nm, AOI = 0°.

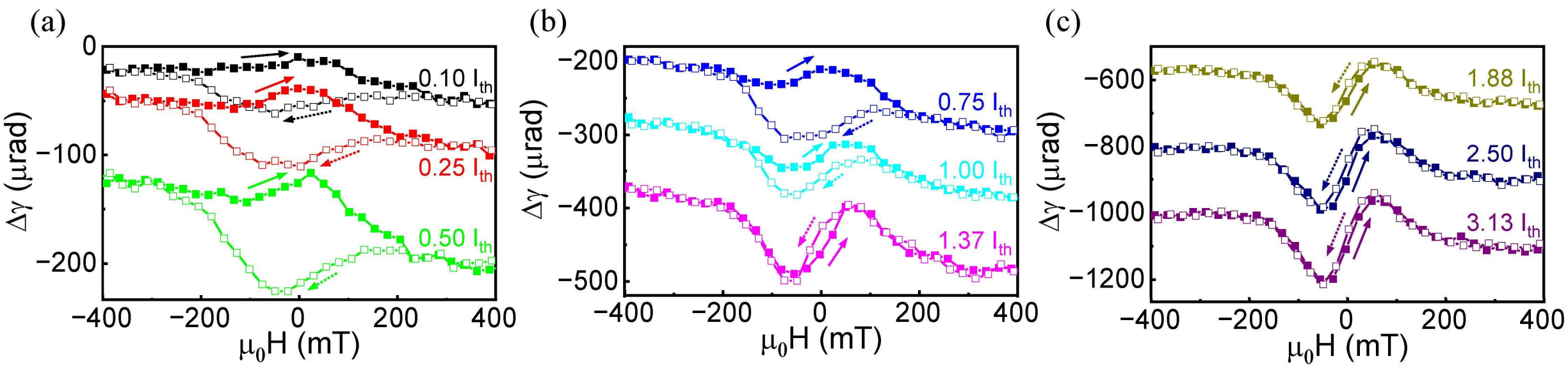

**Fig. S12: Pump-intensity dependence in the concentric geometry. a**, **b**, and **c**, Magnetic-field-dependent ellipticity hysteresis loops measured for $\Delta t = 20$ ps at different pump intensities. Pump wavelength 800 nm, probe wavelength 532 nm.

Figure S12 shows how the magnetic-field-dependent ellipticity loops evolve with pump intensity in the concentric geometry. At the lowest intensity of 0.1 $I_{th}$ , the hysteresis loop closely resembles the static loops, which are a monotonic function of the external field at all temperatures (see Figure 2d). However, as the intensity increases, the character of the loops changes considerably. In particular, as the laser fluence approaches the threshold fluence $I_{th}$, the MO signals become a strongly nonmonotonic function of the applied magnetic field. Moreover, the coercive fields are strongly reduced above $I_{th}$ (Figure 6f). To express the change in loop character, Figure 6d shows the amplitude of the hysteresis loops, with the sign defined as follows when changing the field from its negative extreme. For monotonic functions of the external field, resembling the static case (Figure 2d), the sign of the loops is defined as positive. In contrast, when the change in the MO signal while reducing the field to 0 mT exceeds the loop amplitude, the amplitude is defined as negative.

Figure S13 shows the pump-intensity dependence of the $\Delta\gamma^{dif}$ signal in the concentric geometry, measured for a magnetic field of 16 mT, over a long time range up to 3 ns. Importantly, the decay of this signal does not change above $I_{th}$, resembling the dynamics of $\Delta T/T$.

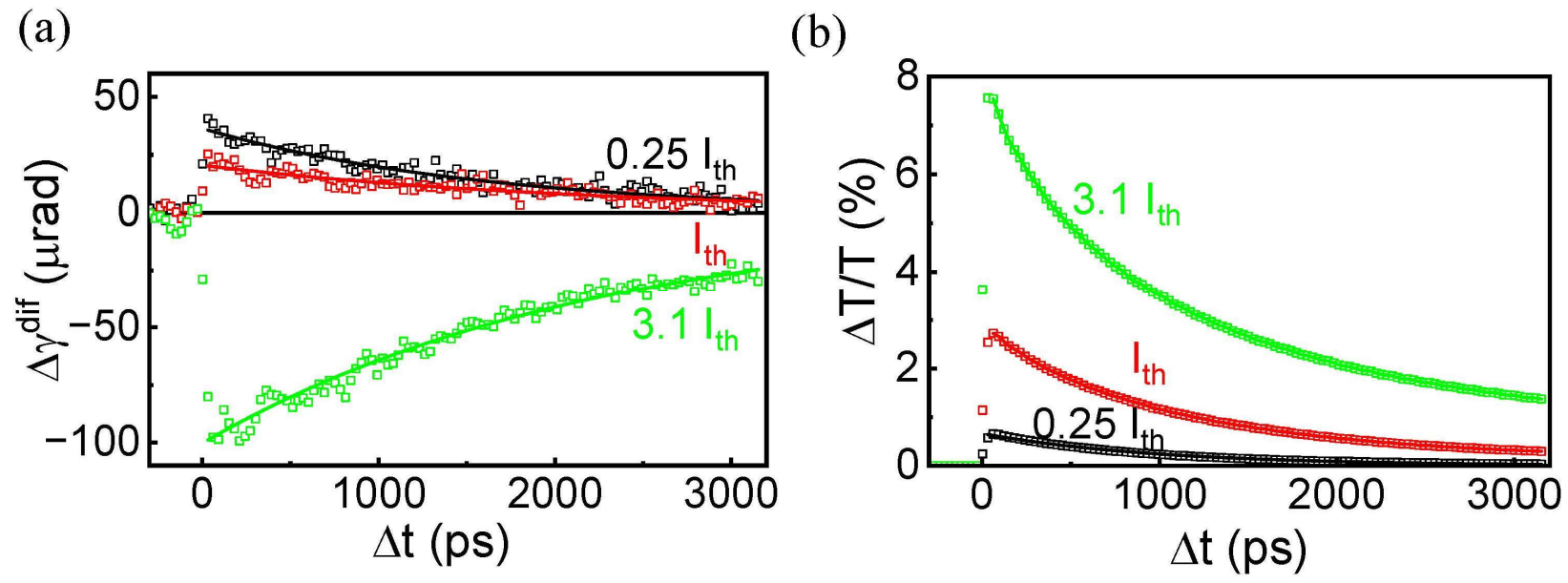

**Fig. S13: Long time range dynamics measured in the concentric geometry. a,** Ellipticity $\Delta\gamma^{dif}$ signal measured as a function of time delay for different pump intensities (points), lines are the fits using a monoexponential decay function plus background with the time constant of 1 ns. **b**, Dynamics of transient transmission $\Delta T/T$ measured for the same pump intensities (points), lines are the fits using a monoexponential decay function plus background with the time constant of 1 ns. Pump wavelength 800 nm, probe wavelength 532 nm.

## Supplementary Note 3. Comparison of results measured in (Ge,Mn)Te with results reported for (Ga,Mn)As

In this Supplementary Note, we compare in detail the pump-induced MO signals observed in thin films of (Ge,Mn)Te with results reported for thin films of (Ga,Mn)As, which is the most thoroughly studied diluted magnetic semiconductor [S3].

In semiconductors, absorption of an intense femtosecond laser pulse photoexcites populations of excess electrons and holes, which thermalize within ≈ 100 fs through various scattering processes, forming local Fermi-Dirac distributions in the conduction and valence bands, respectively. Because the pump photon energy used in our experiment, 1.24–3.1 eV, is much larger than the indirect band gap of (Ge,Mn)Te (0.85 eV in GeTe [S4]), the excess energy of the photoexcited carriers is dissipated to the lattice by electron–phonon and hole–phonon scattering processes on a (sub)picosecond time scale. This creates an excess concentration of electrons and holes in the material. Since both (Ga,Mn)As [S3] and (Ge,Mn)Te [S5] are *p*-type semiconductors, the lifetime of photoexcited free electrons is much shorter than that of photoexcited free holes because traps for holes in the material are already filled by the background holes. When the photoexcited electrons are trapped at defect states, they transfer their energy to the lattice, leading to a temperature increase. Non-radiative recombination of photoexcited free holes with the trapped electrons increases the lattice temperature even further. Finally, the increased lattice temperature of the photoexcited semiconductor film relaxes back to its base temperature by heat diffusion to the sample substrate.

Unlike in (Ga,Mn)As [S6–S8], we used pump photon energies much larger than the band gap energy of (Ge,Mn)Te, resulting in high excess energies of the photoexcited carriers. These excess energies dissipate into the lattice through carrier–phonon scattering processes on a (sub)picosecond timescale. In our experiments, the pump wavelength was varied from 1000 nm (1.24 eV) to 400 nm (3.1 eV). Although the total amount of dissipated heat changes substantially between these wavelengths, the measured MO signals do not change significantly, as shown in Supplementary Figure S8 (see also Figure 5f). This rules out excess-energy dissipation–induced lattice heating as the dominant mechanism leading to magnetization precession in (Ge,Mn)Te. Consequently, trapping of free photoelectrons by defect centers is the fastest effect leading to a transient increase in the lattice temperature, which can, in principle, be responsible for the observed magnetization precession. This temperature increase builds up on a timescale of ≈ 10 ps and persists for ≈ 1 000 ps in (Ga,Mn)As [S7], and similar timescales are observed in the GeMnTe sample as well; see Supplementary Figure S9c and Figure S13 for experimental data illustrating the initial stages of the carrier dynamics and the signal decay over a nanosecond time range, respectively. As the magnetism in (Ge,Mn)Te is hole-mediated, the other effect, which can lead to pump-induced changes in the magnetic properties, is the excess density of photoholes in the valence band. There are several approaches by which these two contributions can be identified.

The first type of experimental data, in which thermal and nonthermal excitation mechanisms can be distinguished experimentally, is pump-induced precession of magnetization, which has been studied extensively in $Ga_{1-x}Mn_xAs$ [S6–S8]. In this material, thermally induced precession of magnetization is observed in samples with low Mn doping ($x$ < 7%; see Figure 5 in Ref. S7) at relatively low pump fluences ($I$ < 70 μJ cm$^{-2}$ for $x$ = 3%; see Figure 2 in Ref. S7), where the position of the easy axis is determined by competing biaxial and uniaxial magnetic anisotropies that scale with magnetization as $M^4$ and $M^2$, respectively [S7,S8]. When non-radiative recombination of photoelectrons produces a transient increase in the lattice temperature, $M$ is reduced and, consequently, the uniaxial anisotropy is enhanced relative to the biaxial anisotropy. This leads to a transient tilt of the easy axis orientation in the material, around which the Mn moments start to precess [S7]. The key fingerprint of this mechanism is the relatively slow rise time of the MO signal, which develops over tens of picoseconds (see Figure 2 in Ref. S7) due to lattice heating resulting from photoelectron trapping [S7].

For higher Mn doping or higher fluences, the spin-orbit torque due to excess hole density in the spin-orbit-coupled, exchange-split valence band becomes the dominant excitation mechanism [S7]. The increased number of occupied hole states (compared with the equilibrium state in the dark) can generate a nonequilibrium hole spin polarization that is misaligned with the equilibrium orientation of the Mn moments. This non-equilibrium photohole polarization persists over the timescale of hole recombination, during which it exerts a torque on the Mn local moments. Experimentally, it appears as an impulsive tilt of the magnetization, which makes it possible to distinguish the spin-orbit torque from the considerably slower thermal excitation mechanism (see Figure 2 in Ref. S7).

In (Ge,Mn)Te, the oscillatory MO signal is fully developed in less than $\approx$ 1 ps after absorption of the pump pulse (see the inset in Figure 5c and Supplementary Note 2.8 for a detailed discussion). Therefore, the transient temperature increase on a timescale of $\approx$ 10 ps (Supplementary Figure S9c) cannot cause the magnetization precession in the (Ge,Mn)Te sample. The rapid energy transfer from photoexcited carriers to phonons (and magnons) also generates thermal stress in the material, which can launch coherent acoustic phonon wave packets (strain waves) and induce magnetization precession [S9]. However, because the duration of the induced strain pulses is on the order of tens of picoseconds (see Figure 1b in Ref. S9), this mechanism is again not compatible with the ultrafast process that triggers magnetization precession on the sub-picosecond timescale. This leaves the pump-induced excess hole density in the valence band, which creates a non-equilibrium hole spin polarization misaligned with the equilibrium orientation of the Mn moments and acts as a spin-orbit torque [S7], as the most probable cause of the observed impulsive tilt of the magnetization, leading to magnetization precession in the (Ge,Mn)Te sample.

The second type of experimental data, in which thermal and non-thermal excitation mechanisms can be distinguished experimentally, is the pump-induced change in hysteresis loops. A pump-induced temperature increase suppresses magnetic ordering in the ferromagnet (i.e., demagnetization), resulting in a reduction of its MO response [S10]. In contrast, a pump-induced enhancement of the MO signal was observed in the (Ge,Mn)Te sample, as evident from the identical sign of the measured hysteresis loops in the static experiment (Figure 2d) and in the low-intensity dynamical data (Figure 1e and Figure 6d). This MO signal enhancement can be attributed to an increase in hole concentration, which strengthens the hole-mediated magnetism in (Ge,Mn)Te [S5], as observed in (Ga,Mn)As [S11]. Moreover, the magnetic-field dependence of the MO signals becomes strongly non-monotonic for laser fluences approaching $I_{th}$ (see Supplementary Figure S12), whereas the static hysteresis loops remain unchanged (i.e., monotonic functions of the external field) at all temperatures (Figure 2d). We attribute this effect, which is absent in (Ga,Mn)As, to the action of spin–orbit torque associated with the laser-induced change in the Rashba spin–orbit interaction, leading to a reconfiguration of the correlated spin-glass structure of the GeMnTe sample.

## Supplementary references